\documentclass{article}
\usepackage[utf8]{inputenc}
\usepackage{geometry}
\usepackage{amssymb}
\usepackage{amsmath}
\usepackage{dsfont}
\usepackage{tabularx}
\usepackage{siunitx}
\usepackage{subfig}
\usepackage{booktabs}
\usepackage{multirow}
\usepackage{tikz}
\usepackage{numprint}
\usetikzlibrary{positioning}
\usepackage{graphicx}
\graphicspath{ {./figures/} }
\usepackage[backend=bibtex,giveninits=true,sorting=nty, bibstyle=authoryear, citestyle=authoryear, maxnames=3,natbib]{biblatex}
\usepackage[hidelinks]{hyperref}
\usepackage{xcolor}
\usepackage{authblk}

\title{HARNet: A Convolutional Neural Network for Realized Volatility Forecasting}
\author[$\dagger$,$\ddagger$,*]{Rafael Reisenhofer}
\author[$\ddagger$,$\mathsection$,*]{Xandro Bayer}
\author[$\ddagger$,$\mathsection$]{Nikolaus Hautsch}

\affil[$\dagger$]{Faculty of Mathematics, University of Vienna, Oskar-Morgenstern-Platz 1, A-1090 Vienna, Austria}
\affil[$\ddagger$]{Research Network Data Science @ Uni Vienna, Kolingasse 14-16, A-1090 Vienna, Austria}
\affil[$\mathsection$]{Department of Statistics and Operations Research, University of Vienna, Oskar-Morgenstern-Platz 1, A-1090 Vienna, Austria}
\affil[*]{These authors contributed equally}

\date{}

\newcommand{\conv}{\ast}
\newcommand{\cdconv}[1]{\cconv_{#1}}
\newcommand{\cconv}{\conv^{\mathrm{c}}}
\newcommand{\RV}{RV}
\newcommand{\RS}{RS}
\newcommand{\loss}{\ell}
\newcommand{\LMSE}{\loss_{\mathrm{MSE}}}
\newcommand{\LMAE}{\loss_{\mathrm{MAE}}}
\newcommand{\LQLIKE}{\loss_{\mathrm{Q}}}
\newcommand{\filt}[1]{h^{(#1)}}
\newcommand{\avgf}[1]{\filt{#1}_{\mathrm{avg}}}
\newcommand{\HARPhi}[1]{\Phi^{\mathrm{HAR#1}}_\theta}

\npthousandsep{,}
\newcommand\blfootnote[1]{%
  \begingroup
  \renewcommand\thefootnote{}\footnote{#1}%
  \addtocounter{footnote}{-1}%
  \endgroup
}

\begin{document}

\maketitle
\blfootnote{Email: rafael.reisenhofer@uni-bremen.de, xandro.bayer@univie.ac.at, nikolaus.hautsch@univie.ac.at}
\begin{abstract}
Despite the impressive success of deep neural networks in many application areas, neural network models have so far not been widely adopted in the context of volatility forecasting. In this work, we aim to bridge the conceptual gap between established time series approaches, such as the Heterogeneous Autoregressive (HAR) model \parencite{corsi2009simple}, and state-of-the-art deep neural network models. The newly introduced HARNet is based on a hierarchy of dilated convolutional layers, which facilitates an exponential growth of the receptive field of the model in the number of model parameters. HARNets allow for an explicit initialization scheme such that before optimization, a HARNet yields identical predictions as the respective baseline HAR model. Particularly when considering the QLIKE error as a loss function, we find that this approach significantly stabilizes the optimization of HARNets. We evaluate the performance of HARNets with respect to three different stock market indexes. Based on this evaluation, we formulate clear guidelines for the optimization of HARNets and show that HARNets can substantially improve upon the forecasting accuracy of their respective HAR baseline models. In a qualitative analysis of the filter weights learnt by a HARNet, we report clear patterns regarding the predictive power of past information. Among information from the previous week, yesterday and the day before, yesterday's volatility makes by far the most contribution to today's realized volatility forecast. Moroever, within the previous month, the importance of single weeks diminishes almost linearly when moving further into the past.
\end{abstract}
\section{Introduction}
%
Volatility is of critical importance in risk management, asset pricing and portfolio construction. Over the past years, extensive research efforts have been devoted to the development and evaluation of volatility forecasting models. A substantial body of research focuses on the construction of estimators for daily volatility exploiting high-frequency information. Seminal contributions (see, e.g., \cite{andersen2001distribution,barndorff2002econometric}) provided the foundation for using the realized variance, i.e., the sum of squared intraday returns, as a consistent estimator for the latent integrated variance of an underlying diffusion process. These approaches have been extended and robustified, and built the starting point for an active line of research on the development of high-frequency-based volatility estimators (see, e.g., \cite{ait2014high}). At the same time, these developments triggered the need to develop appropriate time series models to predict realized volatility. One of the most popular and empirically successful models is the heterogeneous autoregressive (HAR) model proposed by \citet{corsi2009simple}, which captures the typically strong persistency of daily realized variances in a simple and efficient way.

%
In the past decade, deep neural networks (DNNs) have revolutionized many areas of machine learning such as image classification, machine translation or speech recognition. The most widely used neural network (NN)-based approaches for forecasting sequential data are recurrent NNs,  such as the so-called long short-term memory (LSTM) architecture \parencite{hochreiter1997long}. Originally developed for image data sets, NNs with convolutional layers -- so-called convolutional neural networks (CNNs) -- have more recently been applied with impressive success for time series prediction in the realm of natural language processing \parencite{oord2016wavenet, kalchbrenner2016neural, dauphin2017language, gehring2016convolutional, gehring2017convolutional}. In the context of financial time series, \citet{miura_crypto_artificial_2019} investigated numerous machine learning approaches, including CNNs, for predicting the realized volatility of Bitcoin returns, while \citet{borovykh2017conditional} employed a dilated convolutional network for forecasting stock index returns. A main advantage of dilated convolutional layers is that a linear increase in depth allows for an exponential increase of the considered time horizon. Such architectures should thus be particularly powerful for tasks, where large time horizons are potentially useful, but machine learning models often suffer from overfitting with a growing number of model parameters.

The applicability of feed-forward and recurrent NN models in the context of realized volatility forecasting has recently been explored by \citet{bucci_realized_2020}, \citet{christensen2021machine} and \citet{poon2020machine}. The fact that so far, NN-based approaches have not been widely adopted in the area of financial econometrics could be explained by the fact that NN models are often used as off-the-shelve black boxes whose results are typically difficult to interpret, in particular in comparison to a simple and well-understood approach like the HAR model. In this work, we aim to bridge the conceptual gap between established methods such as the HAR model and state-of-the-art deep learning-based approaches by introducing so-called 'HARNets' as a novel NN model for forecasting realized volatility.

A HARNet is based on hierarchies of dilated convolutional layers. It is strongly inspired by the well-known HAR model in the sense that each layer of a HARNet computes features with respect to different time horizons, which are comparable to weekly and monthly aggregates as in the original HAR model. In particular, HARNets allow for an explicit initialization scheme for the respective model weights such that before optimization, a HARNet yields identical predictions as the respective baseline HAR model. This approach not only significantly stabilizes the optimization process but also facilitates an in-depth analysis of the performance of HARNets relative to different fits of their respective HAR baselines models.

We briefly summarize important concepts from realized volatility forecasting and the application of CNNs for time series forecasting in Sections~\ref{sec:rvhar}~and~\ref{sec:cnns}, respectively.

The HARNet model as well as our approaches to initializing and optimizing the respective model parameters are presented in Section~\ref{sec:method}.

In Section~\ref{sec:results}, we conduct a detailed evaluation of the performance of HARNets relative to their respective baseline HAR models. In particular, we consider three different loss functions, the mean absolute error (MAE), the mean squared error (MSE) and the QLIKE error. Moreover, we evalute three different estimation approaches for fitting the HAR baseline model, namely ordinary least squares (OLS), weighted least squares (WLS) and $\log$-OLS, where the input time series is logarithmized before being processed by the model. Based on our initial findings, we further address questions regarding the stability of the optimization process and the generalization properties of optimized HARNets, and eventually formulate clear guidelines for the optimization of HARNets. In particular, we also conduct a qualitative analysis of the filter weights learnt by HARNets on the different convolutional layers, which is presented in Section~\ref{sec:interpret}.

In Section~\ref{sec:semivar}, we consider extended HARNets for higher-dimensional input time series, which also contain information regarding the realized semivariance as well as jump variation and again evaluate their performance relative to their extended HAR-like baseline models.  
\section{Forecasting realized volatility with the HAR model}
\label{sec:rvhar}
We assume that the dynamics of the log price of a financial asset evolve according to the stochastic differential equation
\begin{equation}
    dp(t) = \mu(t)dt + \sigma(t)dW(t),
\end{equation}
where the c\`{a}dl\`{a}g finite variation process $\mu(t)$ denotes the drift, $W(t)$ is a standard Brownian motion, and $\sigma(t)$ is the spot volatility, which is independent of $W(t)$. The daily integrated variance
\begin{equation}
    IV_{t} = \int_{t-1}^{t} \sigma^2(\omega)d\omega
\end{equation}
 at a day $t$ is unobservable in practice, even ex-post. However, it has been shown that the daily sum of $N$ intraday squared returns
\begin{equation}
    \label{eq:rv}
    \RV_{t}=\mathop{\sum_{n=0}^{N-1}r_{t,n}^{2}},
\end{equation}
 where $r_{t,n} = p (t-n\Delta) - p(t - (n + 1)\Delta)$ with $\Delta=1/N$, is a consistent estimator for $IV_t$ when $N\to\infty$ \parencite{skeptics}.

The most popular model for forecasting $\RV_{t}$ is the so-called heterogeneous autoregressive model introduced by \citet{corsi2009simple}. The model has a simple autoregressive structure and includes averages of daily realized variances over different time horizons. For an aggregation period $j\in \mathbb{N}$, we denote the respective average of daily realized variances by 
\begin{equation}
\RV_{t}^{(j)} = \frac{1}{j} \sum_{n=0}^{j-1} \RV_{t-n}. 
\end{equation}
\citet{corsi2009simple} proposes using aggregation periods of $1$, $5$, and $22$, corresponding to daily, weekly, and monthly RVs, respectively. The HAR model is then defined as
\begin{equation}
\label{eq:HAR}
\RV_t = \beta_0 + \beta_1 \RV_{t-1} + \beta_2 \RV^{(5)}_{t-1} + \beta_3 \RV^{(22)}_{t-1} + \epsilon_t,
\end{equation}
where the parameters $\beta_0, \ldots, \beta_3$, are usually estimated by ordinary least squares (OLS). 
%
However, since it is well-known that time series of realized variances exhibit properties such as conditional heteroscedasticity and non-Gaussianity, OLS is not ideal. To overcome this, we follow \citet{patton_good_2015} and alternatively estimate the HAR model using  weighted least squares (WLS) with weights chosen as the inverse of the (OLS) fitted value. Another popular approach to deal with heteroscedasticity and non-Gaussianity is to transform the data logarithmically. For the remainder of this paper, this method will be denoted as $\log$-OLS. %
\section{Dilated convolutional NNs for time series forecasting}
\label{sec:cnns}
In this section, we briefly review the basic concepts and definitions underlying dilated convolutional neural networks (NNs) in the context of time series forecasting.

Like the majority of popular NN models, dilated convolutional NNs belong to the general class of feedforward NNs. Such networks transform a high-dimensional input vector, also called the \emph{input layer}, by consecutively applying affine linear mappings and non-linear functions. An affine linear mapping followed by a non-linear function, which acts component-wise on the respective input, is called a \emph{hidden layer}. The layer that yields the final output of the NN is typically called the \emph{output layer}. The single entries of the vectors obtained at the different layers of the NN are called \emph{input neurons}, \emph{hidden neurons}, and \emph{output neurons}, respectively. The number of consecutively applied layers defines the \emph{depth} of the NN, while the number of neurons in a single layer defines the \emph{width} of the NN. The non-linearity usually remains the same across all hidden layers of the network. The parameters that define the affine mappings can be different from layer to layer and are usually referred to as the \emph{weights} of the NN. Note that the terms 'parameters' and 'weights' both refer to the values that define what a given NN model actually computes and are used interchangeably in the machine learning literature. The weights of a NN are \emph{learnt} during \emph{training} in the sense that they are iteratively updated with the goal of minimizing a predefined loss function (cf.~Section~\ref{sec:lossandopt}). Before training, NN weights are usually initialized randomly. Sometimes it can be helpful to keep certain parameters of a NN model fixed throughout training. In this case, one differentiates between \emph{trainable} and \emph{non-trainable} weights. Informally, the extent to which a NN can compute different functions when varying the values of its trainable weights is often called its \emph{capacity}, or \emph{expressivity}. Increasing the depth or the width of a NN model usually also increases its capacity. 

Let $d$ denote the dimension of the input vector $x$ and $L$ the number of layers in a feedforward NN where the $l$-th layer consists of $N_l$ neurons. Each layer is then defined by an affine linear mapping
\begin{equation}
\label{eq:affinemapping}
    A_{l}:\mathbb{R}^{N_{l-1}}\rightarrow\mathbb{R}^{N_{l}}\colon x\mapsto W_lx+b_l,\;\;\;1\leq l\leq L,
\end{equation}
with $N_{0}=d$, weight matrices $W_l\in\mathbb{R}^{N_{l-1}\times N_{l}}$, and bias vectors $b_l\in\mathbb{R}^{N_{l}}$. Let $\sigma$ denote a real-valued non-linear function that acts component-wise on the input. The respective feedforward NN defines a mapping from $\mathbb{R}^d\to\mathbb{R}^{N_L}$, namely 
\begin{equation}
\label{eq:ffwdnn}
\Phi(x)=A_{L}\sigma\left(A_{L-1}\sigma\left(\ldots\sigma\left(A_{1}(x)\right)\right)\right),\;\;\;x\in\mathbb{R}^{d}.
\end{equation}
A convolutional neural network (CNN) is a special case of a feedforward NN where the affine mappings $A_l$ are chosen to define convolution operators. The one-dimensional convolution of a sequence $x$ with a filter $h$ is defined as
\begin{equation}
\label{eq:conv}
 (x\conv h)_t = \sum_{n=-\infty}^{\infty}x_{t - n}h_n,\;\;\;x,h\in\mathbb{R}^{\mathbb{Z}}. 
\end{equation}
CNNs first rose to prominence when deep CNNs revolutionized the field of image classification by outperforming all previously existing approaches by significant margins \parencite{krizhevsky2012imagenet}. The basic idea behind deep CNNs is that sliding a filter over a given input is a simple means of detecting local features. In the context of image classification, these features are usually certain types of edges or corners in early layers but can become increasingly complex in deeper layers where filters are sensitive to certain geometrical shapes or even objects. A main advantage of CNNs is that only considering convolution operators imposes a strong structural constraint on the respective linear mapping $A_l$, which drastically reduces the number of model parameters when working with high-dimensional input data such as images.

Deep CNNs have also been widely applied for time series prediction \parencite{oord2016wavenet, borovykh2017conditional}. When working with time series, it is usually required that a filter can only aggregate information from the past and present, but not from the future. This type of convolution is known as \emph{causal} convolution and can be written as
\begin{equation}
\label{eq:cconv}
 (x\cconv h)_t = \sum_{n=0}^{N-1}x_{t - n}h_{n},\;\;\;x\in\mathbb{R}^{\mathbb{Z}}, h\in\mathbb{R}^N,
\end{equation}
where $h$ denotes a finite filter of length $N$.

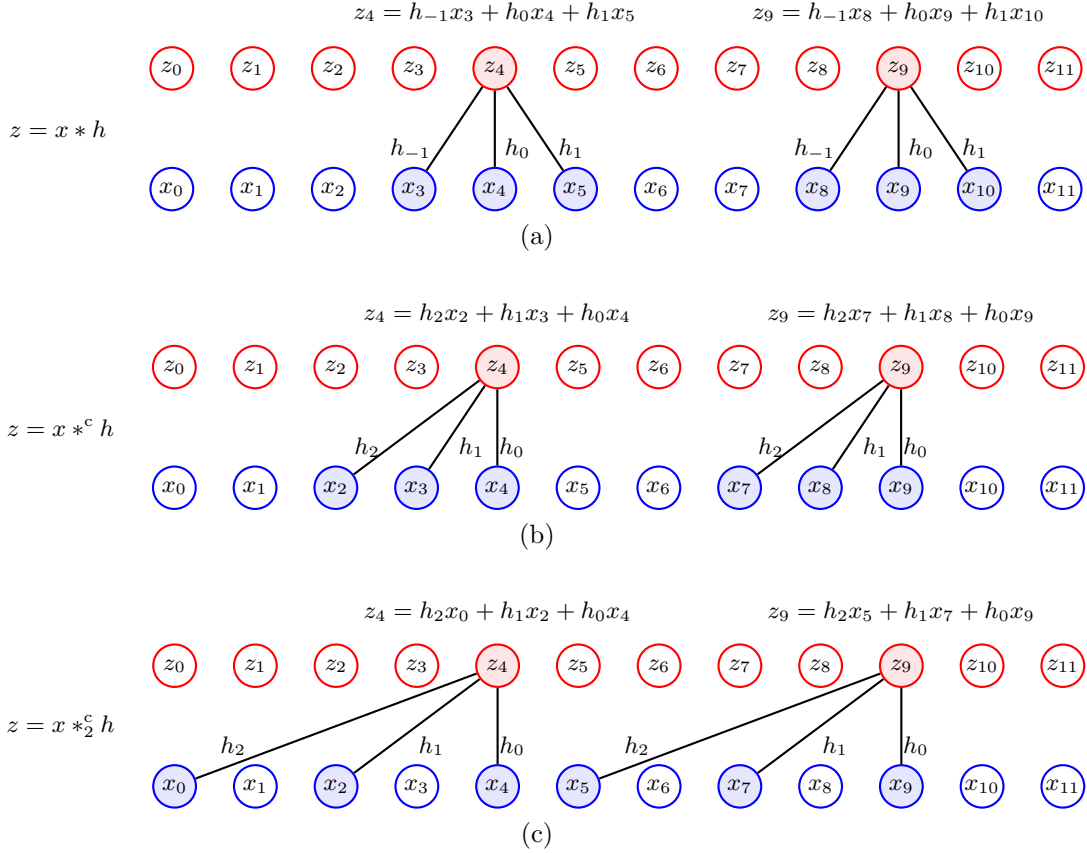
\begin{figure}[tb!]
\centering
\subfloat[]{
\begin{tikzpicture}[x=\textwidth,y=2cm,
tsblue/.style={thick,draw=blue,circle,minimum size=16, inner sep=0pt},
tsred/.style={thick,draw=red,circle,minimum size=16, inner sep=0pt},
tsbluefill/.style={thick,draw=blue,fill=blue!10,circle,minimum size=16, inner sep=0pt},
tsredfill/.style={thick,draw=red,fill=red!10,circle,minimum size=16, inner sep=0pt}]
  \tikzstyle{every node}=[font=\small]
  \def\N{12}
  \pgfmathparse{\N-1}
  \node (desc) at (0,0.6) {$z = x\conv h$};
  \foreach \i [evaluate={\x=0.1 + 0.85*\i/\N; \y=1.0}]in {0,...,\pgfmathresult}{ 
    \ifthenelse{\i = 4\OR \i = 9}{
      \node[tsredfill] (z-\i) at (\x,\y) {$z_{\i}$};
    }{
      \node[tsred] (z-\i) at (\x,\y) {$z_{\i}$};
    }
  }
  \foreach \i [evaluate={\x=0.1 + 0.85*\i/\N; \y=0.2}]in {0,...,\pgfmathresult}{ 
        \ifthenelse{\(\i > 2 \AND \i < 6\) \OR \(\i > 7 \AND \i < 11\)}{
      \node[tsbluefill] (x-\i) at (\x,\y) {$x_{\i}$};
    }{
      \node[tsblue] (x-\i) at (\x,\y) {$x_{\i}$};
    }
  }
  \foreach \iz in {4,9}{
    \foreach \iy [evaluate={\ix=int(\iz+\iy);}] in {-1}{
      \draw[thick] (x-\ix) -- node[below left= 0pt and 5pt] {$h_{\iy}$} (z-\iz);
    }
    \foreach \iy [evaluate={\ix=int(\iz+\iy);}] in {1}{
      \draw[thick] (x-\ix) -- node[below right = 0pt and 5pt] {$h_{\iy}$} (z-\iz);
    }
    \foreach \iy [evaluate={\ix=int(\iz+\iy);}] in {0}{
      \draw[thick] (x-\ix) -- node[below right] {$h_{\iy}$} (z-\iz);
    }
  }
  \node [above = 5pt of z-4] (zlbl-4) {$z_{4} = h_{-1}x_3 + h_{0}x_4 + h_{1}x_{5}$};
  \node [above = 5pt of z-9] (zlbl-9) {$z_{9} = h_{-1}x_8 + h_{0}x_9 + h_{1}x_{10}$};
\end{tikzpicture}}\\[.5cm]

\subfloat[]{
\begin{tikzpicture}[x=\textwidth,y=2cm,
tsblue/.style={thick,draw=blue,circle,minimum size=16, inner sep=0pt},
tsred/.style={thick,draw=red,circle,minimum size=16, inner sep=0pt},
tsbluefill/.style={thick,draw=blue,fill=blue!10,circle,minimum size=16, inner sep=0pt},
tsredfill/.style={thick,draw=red,fill=red!10,circle,minimum size=16, inner sep=0pt}]
  \tikzstyle{every node}=[font=\small]
  \def\N{12}
  \pgfmathparse{\N-1}
  \node (desc) at (0,0.6) {$z = x\cconv h$};
  \foreach \i [evaluate={\x=0.1 + 0.85*\i/\N; \y=1.0}]in {0,...,\pgfmathresult}{ 
    \ifthenelse{\i = 4\OR \i = 9}{
      \node[tsredfill] (z-\i) at (\x,\y) {$z_{\i}$};
    }{
      \node[tsred] (z-\i) at (\x,\y) {$z_{\i}$};
    }
  }
  \foreach \i [evaluate={\x=0.1 + 0.85*\i/\N; \y=0.2}]in {0,...,\pgfmathresult}{ 
        \ifthenelse{\(\i > 1 \AND \i < 5\) \OR \(\i > 6 \AND \i < 10\)}{
      \node[tsbluefill] (x-\i) at (\x,\y) {$x_{\i}$};
    }{
      \node[tsblue] (x-\i) at (\x,\y) {$x_{\i}$};
    }
  }
  \foreach \iz in {4,9}{
    \foreach \iy [evaluate={\ix=int(\iz-\iy);}] in {2}{
      \draw[thick] (x-\ix) -- node[below left = 0pt and 10pt] {$h_{\iy}$} (z-\iz);
    }
    \foreach \iy [evaluate={\ix=int(\iz-\iy);}] in {1,0}{
      \draw[thick] (x-\ix) -- node[below right = 0pt and -3pt] {$h_{\iy}$} (z-\iz);
    }
  }
  \node [above = 5pt of z-4] (zlbl-4) {$z_{4} = h_{2}x_2 + h_{1}x_3 + h_{0}x_{4}$};
  \node [above = 5pt of z-9] (zlbl-9) {$z_{9} = h_{2}x_7 + h_{1}x_8 + h_{0}x_{9}$};
\end{tikzpicture}}\\[.5cm]

\subfloat[]{
\begin{tikzpicture}[x=\textwidth,y=2cm,
tsblue/.style={thick,draw=blue,circle,minimum size=16, inner sep=0pt},
tsred/.style={thick,draw=red,circle,minimum size=16, inner sep=0pt},
tsbluefill/.style={thick,draw=blue,fill=blue!10,circle,minimum size=16, inner sep=0pt},
tsredfill/.style={thick,draw=red,fill=red!10,circle,minimum size=16, inner sep=0pt}]
  \tikzstyle{every node}=[font=\small]
  \def\N{12}
  \pgfmathparse{\N-1}
  \node (desc) at (0,0.6) {$z = x\cdconv{2} h$};
  \foreach \i [evaluate={\x=0.1 + 0.85*\i/\N; \y=1.0}]in {0,...,\pgfmathresult}{ 
    \ifthenelse{\i = 4\OR \i = 9}{
      \node[tsredfill] (z-\i) at (\x,\y) {$z_{\i}$};
    }{
      \node[tsred] (z-\i) at (\x,\y) {$z_{\i}$};
    }
  }
  \foreach \i [evaluate={\x=0.1 + 0.85*\i/\N; \y=0.2}]in {0,...,\pgfmathresult}{ 
        \ifthenelse{\i= 0 \OR \i= 2 \OR \i= 4 \OR \i= 5 \OR \i= 7 \OR \i= 9}{
      \node[tsbluefill] (x-\i) at (\x,\y) {$x_{\i}$};
    }{
      \node[tsblue] (x-\i) at (\x,\y) {$x_{\i}$};
    }
  }
  \foreach \iz in {4,9}{
    \foreach \iy [evaluate={\ix=int(\iz-2*\iy);}] in {2}{
      \draw[thick] (x-\ix) -- node[below left = 0pt and 30pt] {$h_{\iy}$} (z-\iz);
    }
    \foreach \iy [evaluate={\ix=int(\iz-2*\iy);}] in {1,0}{
      \draw[thick] (x-\ix) -- node[below right = 0pt and -3pt] {$h_{\iy}$} (z-\iz);
    }
  }
  \node [above = 5pt of z-4] (zlbl-4) {$z_{4} = h_{2}x_0 + h_{1}x_2 + h_{0}x_{4}$};
  \node [above = 5pt of z-9] (zlbl-9) {$z_{9} = h_{2}x_5 + h_{1}x_7 + h_{0}x_{9}$};
\end{tikzpicture}}
\caption{Three different types of one-dimensional convolutions. (a) One-dimensional convolution of a time series $x$ with a filter $h$. (b) Causal convolution of a time series $x$ with a filter $h$. (c) Dilated causal convolution of a time series $x$ with a filter $h$.}
\label{fig:1dconv}
\end{figure}

\begin{figure}[tb!]
    \centering
    \includegraphics[width = .99\textwidth]{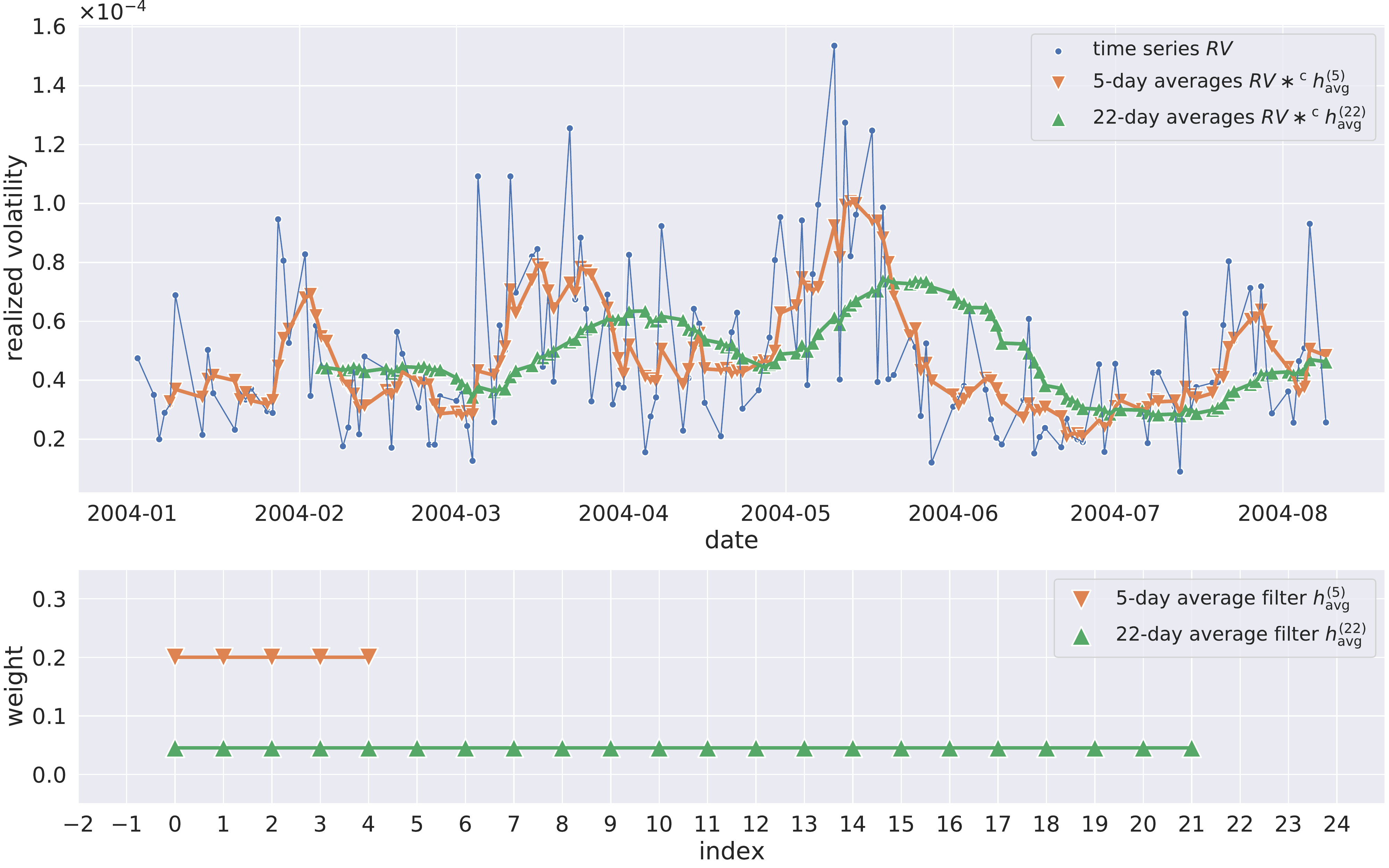}
    \caption{Causal convolution of a realized volatility time series with 5-day and 22-day average filters. The filters $\avgf{5}$ and $\avgf{22}$ are defined in equation~\eqref{eq:avgfilters}.}
    \label{fig:avgfilters}
\end{figure}

The \emph{receptive field} of a model describes the part of an input time series which is actually used by the model to compute a prediction for future values of the time series. That is, the receptive field determines how far a model can see into the past. When only considering convolutions of the form \eqref{eq:cconv}, the length of the receptive field can only grow linearly with the number of filter coefficients, that is, the number of trainable parameters in the model. In situations with long-range dependencies, this can lead to inefficient training and overfitting due to the high number of trainable parameters. A possible remedy is to use compositions of \emph{dilated} convolutions. In a dilated convolution, the inner product with the convolution filter is not based on the consecutive entries of the time series but on entries that are a fixed number of steps apart from each other. Formally, a dilated causal convolution with dilation factor $k$ is defined as 
\begin{equation}
\label{eq:dilconv}
(x \cdconv{k} h)_t = \sum_{n=0}^{N-1}x_{t - kn}h_{n},\;\;\;x\in\mathbb{R}^{\mathbb{Z}}, h\in\mathbb{R}^N,
\end{equation}
where $h$ denotes a finite filter of length $N$. By composing dilated convolutional layers that have exponentially increasing dilation factors, it is possible to cover receptive fields that grow exponentially with the number of trainable model parameters. All three types of convolutions presented in this section are illustrated in Figure~\ref{fig:1dconv}. Examples for causal convolutions of an RV time series with different average filters are shown in Figure~\ref{fig:avgfilters}.
\subsection{Optimization of NNs}
Let $\Phi_\theta\colon \mathbb{R}^d\to \mathbb{R}^{N_L}$ be the function defined by a NN model as in equation~\eqref{eq:ffwdnn}, where $\theta$ denotes the set of all trainable weights. That is, $\theta$ contains all parameters of the matrices $W_l$ and bias vectors $b_l$ that are learnt during training. Here, we consider the case of \emph{supervised learning}. This means that the model is optimized such that its output best matches input-output pairs from a carefully annotated and selected \emph{training data set}. The outputs within the training data set are usually referred to as \emph{labels}. When training a NN architecture to yield one-day-ahead forecasts for RV, the input $x$ in a single input-label pair from the training set typically consists of the complete training RV time series up to a day $t$, while the corresponding label $y$ is chosen as the RV on day $t+1$, that is, $x = \left(\ldots, RV_{t-1}, RV_{t}\right)^{T}$ and $y = RV_{t+1}$. 

For a sequence $Z = \left(x^{(n)}, y^{(n)}\right)_{n=1}^{N}$ of $N$ input-label pairs, we define the \emph{empirical risk} via
\begin{equation}
    \label{eq:emprisk}
    J\left(\theta, Z\right) = \frac{1}{N}\sum_{n=1}^{N}\ell(\Phi_\theta(x^{(n)}), y^{(n)}),
\end{equation}
where $\ell$ is a real-valued loss function. In the context of this work, the labels $y^{(n)}$ are always the consecutive values of the RV time series $x^{(n)}$, which are both taken from a predefined training time series. $\Phi_\theta(x^{(n)})$ represents the forecast of the model with parameters $\theta$.

The sequence $Z$, which can contain the complete training data set, or only selected samples from different parts of the full training time series, is usually referred to as a \emph{batch}. The length $N$ of the sequence $Z$ is called the \emph{batch size}.

The weights $\theta$ are updated iteratively via gradient descent. In its most basic form, the gradient descent update rule can be written as
\begin{equation}
    \label{eq:graddesc}
    \theta^{(i+1)} = \theta^{(i)} - \alpha \nabla_{\theta} J(\theta^{(i)}, Z^{(i)}), 
\end{equation}
where $\alpha > 0$ denotes the learning rate, $Z^{(i)}$ is the training data batch used for the $i$-th update of the parameters $\theta$, and the gradient vector is defined as
\begin{equation}
    \renewcommand*{\arraystretch}{2} 
    \nabla_{\theta} J(\theta^{(i)}, Z^{(i)}) = \begin{pmatrix}\frac{\partial J \left(\theta^{(i)}, Z^{(i)}\right)}{\partial \theta_1}\\\frac{\partial J \left(\theta^{(i)}, Z^{(i)}\right)}{\partial \theta_2}\\\vdots\end{pmatrix}.
\end{equation}

Two of the most popular first-order gradient descent algorithms are ADAM \parencite{adam} and RMSProp \parencite{hinton2012neural}, which both implement adaptive schemes for the learning rate $\alpha$. The gradient vector $\nabla_{\theta} J(\theta^{(i)}, Z^{(i)})$ is usually obtained via backpropagation \parencite{backprop}.

In practice, each batch $Z^{(i)}$ usually contains only a small subset of the complete training data, which is selected randomly for each gradient descent update of the form \eqref{eq:graddesc}. This approach is known as mini-batch stochastic gradient descent (SGD). When only using a single input-label pair in each gradient descent step, the learning algorithm often has difficulties identifying patterns in the data, in particular when the training data is very noisy. Using the entire training data as a single batch, on the other hand, often leads to overfitting. Mini-batch SGD can be seen as a compromise between these two extremes which has been found to be highly applicable in practice. 

Parameters such as the (initial) learning rate $\alpha$, the batch size, the stopping criteria for the learning algorithm, but also the width and depth and the NN are called \emph{hyperparameters}. Hyperparameters define important characteristics of the architecture and the applied learning algorithm which are not changed during training. 

For a more comprehensive overview on deep learning, see for example \citet{deeplearningbook}.
\section{Method}
\label{sec:method}
We aim to exploit the close relationship between the original HAR model (cf. equation~\eqref{eq:HAR}) and NN models that are based on layers of dilated causal convolutions of the form \eqref{eq:dilconv}. In particular, we will consider a NN network model whose weights can be initialized such that, before optimization, it exactly replicates the predictions of a fitted HAR model. This initialization approach significantly stabilizes the optimization process and allows us to perform a rigorous evaluation of the nonlinear NN model relative to the HAR model with respect to different types of data as well as different optimization settings.

We first note that the HAR model can easily be written in terms of convolutions with average filters. Let
\begin{equation}
\label{eq:avgfilters}
    \avgf{j} = \left(\frac{1}{j},\ldots, \frac{1}{j}\right) \in \mathbb{R}^j
\end{equation}
denote an average filter of length $j\in\mathbb{N}$. Then, the averages of daily realized volatility are defined by $\RV_{t}^{(j)} =  \left(\RV\cconv \avgf{j}\right)_{t}$,
and thus, the HAR model \eqref{eq:HAR} can be written as
\begin{equation}
    \label{eq:HARconv}
\RV_t = \left(\beta_0 + \beta_1 \RV\cconv \avgf{1} + \beta_2 \RV\cconv \avgf{5} + \beta_3 \RV\cconv \avgf{22}\right)_{t-1} + \epsilon.
\end{equation}

For the model defined in equation~\eqref{eq:HARconv}, the weights of the filters $\avgf{1}$, $\avgf{5}$, and $\avgf{22}$ are fixed to obtain the respective 1-, 5-, and 22-day averages (cf.~Figure~\ref{fig:avgfilters}). A straightforward but impractical approach to define a CNN with learnable filters based on this model would be to consider a network with a single convolutional layer, which consists of three different filters of length 1, 5, and 22, respectively, with trainable weights, and a linear output layer, which aggregates all filter outputs with trainable weights $\beta_0, \ldots, \beta_3$. With the additional non-linearity, which is applied to all filter outputs, and a total number of 32 trainable parameters ($\beta_0, \ldots, \beta_3$ and the weights of the three filters in the convolutional layer), this model would be significantly more expressive than the HAR model, which has only four parameters. Moreover, such a shallow model with only one hidden layer does not exploit the hierarchical structure of deep CNNs while having the same receptive field as the original HAR model with eight times as many model parameters. As a consequence, this model would be extremely prone to overfitting.

An approach that keeps the number of model parameters low while also increasing the expressiveness of the HAR model such that meaningful filters for different time horizons can be learnt is to consider dilated convolutions and multiplicatively nested time horizons. The latter means that each aggregation period in a sequence of time horizons is an integer multiple of its predecessor (cf.~equation~\ref{eq:multnest}). In particular, this approach yields receptive fields that can grow exponentially in the number of model parameters.

Let $(j_1,\ldots, j_L)$ be an increasing sequence of $L$ aggregation periods, where each period is an integer multiple of its predecessor, that is,
\begin{equation}
\label{eq:multnest}
\exists\; c_l\in\mathbb{N}\colon\; c_lj_l = j_{l+1}\;\;\; \forall\; l\in\{1,\ldots, L-1\}.
\end{equation} Then, the respective averages of daily RVs can be defined recursively via
\begin{equation}
\label{eq:rvavgrec}
\RV^{(j_l)} = \begin{cases}\RV\cconv \avgf{j_1} & \text{if } l = 1,\\\RV^{(j_{l-1})}\cdconv{j_{l-1}} \avgf{j_l/j_{l-1}} & \text{else.} \end{cases}
\end{equation}
%
For the multiplicatively nested sequence of time horizons $(1, 5, 20)$, which closely resembles the sequence of time horizons considered in the original HAR model, this approach would yield a hierarchical model of dilated convolutions with three filters of lengths $1$, $5$, and $4$, respectively, where the second filter computes weekly weighted averages and the third filter monthly weighted averages. The respective model, however, would only contain 14 trainable parameters.
\subsection{HARNet model}
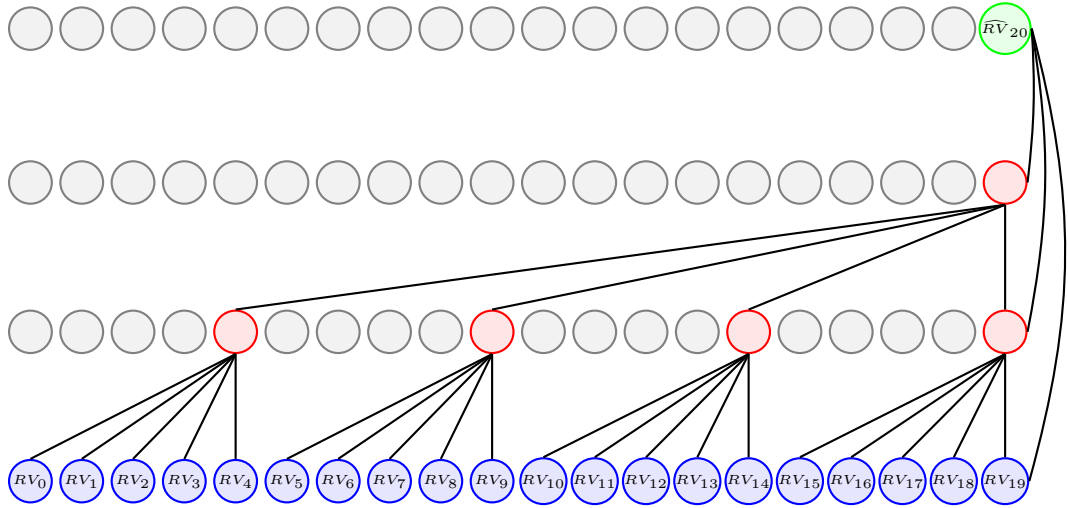
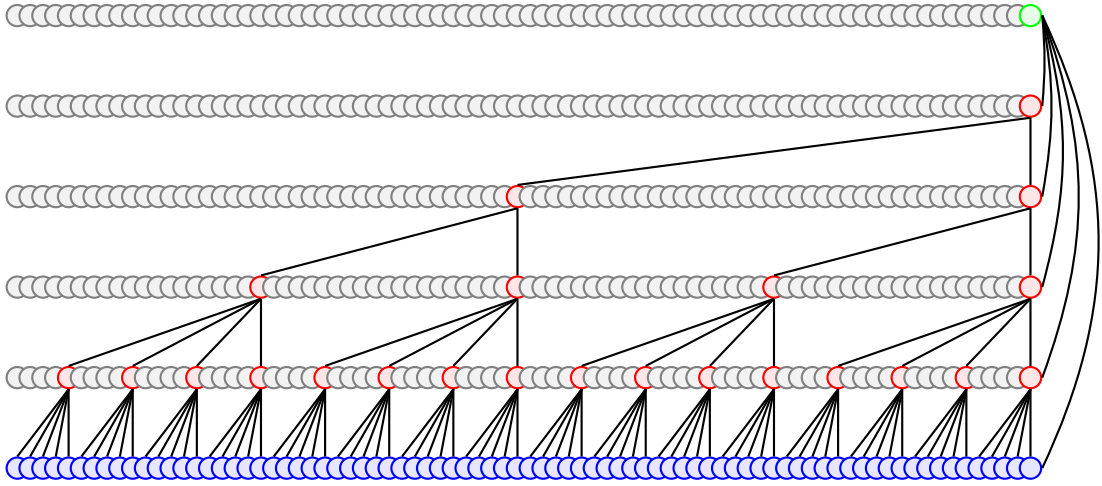
\begin{figure}[th!]
\centering
\subfloat[Network topology of $\HARPhi{20}$ with aggregation periods $(1, 5, 20)$.]{\begin{tikzpicture}[x=\textwidth,y=6cm,
tsbluefill/.style={thick,draw=blue,fill=blue!10,circle,minimum size=16, inner sep=0pt},
tsredfill/.style={thick,draw=red,fill=red!10,circle,minimum size=16, inner sep=0pt},
tsgrayfill/.style={thick,draw=gray,fill=gray!10,circle,minimum size=16, inner sep=0pt},
tsgreenfill/.style={thick,draw=green,fill=green!10,circle,minimum size=16, inner sep=0pt}]
  \tikzstyle{every node}=[font=\tiny]
  \def\N{20}
  \pgfmathparse{\N-1}
  \foreach \i [evaluate={\x=.9*\i/\N; \y=0.0}]in {0,...,\pgfmathresult}{ 
      \node[tsbluefill] (rv-\i) at (\x,\y) {$\RV_{\i}$};
  }
  \foreach \i [evaluate={\x=.9*\i/\N; \y=0.33}]in {0,...,\pgfmathresult}{ 
    \ifthenelse{\i = 4 \OR \i = 9 \OR \i = 14 \OR \i = 19}{
      \node[tsredfill] (rv5-\i) at (\x,\y) {};
    }{
      \node[tsgrayfill] (rv5-\i) at (\x,\y) {};
    }
  }
  
  \foreach \i [evaluate={\x=.9*\i/\N; \y=0.66}]in {0,...,\pgfmathresult}{ 
    \ifthenelse{\i = 19}{
      \node[tsredfill] (rv20-\i) at (\x,\y) {};
    }{
      \node[tsgrayfill] (rv20-\i) at (\x,\y) {};
    }
  }
  
  \foreach \i [evaluate={\x=.9*\i/\N; \y=1.0;\inext = int(\i + 1)}]in {0,...,\pgfmathresult}{ 
    \ifthenelse{\i = 19}{
      \node[tsgreenfill] (pred-\i) at (\x,\y) {$\widehat{\RV}_{\inext}$};
    }{
      \node[tsgrayfill] (pred-\i) at (\x,\y) {};
    }
  }
  
  \foreach \iz in {4,9, 14, 19}{
    \foreach \iy [evaluate={\ix=int(\iz+\iy);}] in {-4,-3,-2,-1,0}{
      \draw[thick] (rv-\ix.north) -- node[left] {} (rv5-\iz.south);
    }
  }
  \foreach \iz in {19}{
    \foreach \iy [evaluate={\ix=int(\iz+5*\iy);}] in {-3,-2,-1,0}{
      \draw[thick] (rv5-\ix.north) -- node[left] {} (rv20-\iz.south);
    }
  }
  \foreach \i in {19}{
    \draw[-,thick] (rv-\i.east) to [bend right=15] node[right] {} (pred-\i.east);
    \draw[-,thick] (rv5-\i.east) to [bend right=10] node[right] {} (pred-\i.east);
    \draw[-,thick] (rv20-\i.east) to [bend right=5] node[right] {} (pred-\i.east);
  }
\end{tikzpicture}}\\[.5cm]

\subfloat[Network topology of $\HARPhi{80}$ with aggregation periods $(1, 5, 20, 40, 80)$.]{\begin{tikzpicture}[x=\textwidth,y=6cm,
tsbluefill/.style={thick,draw=blue,fill=blue!10,circle,minimum size=8, inner sep=0pt},
tsredfill/.style={thick,draw=red,fill=red!10,circle,minimum size=8, inner sep=0pt},
tsgrayfill/.style={thick,draw=gray,fill=gray!10,circle,minimum size=8, inner sep=0pt},
tsgreenfill/.style={thick,draw=green,fill=green!10,circle,minimum size=8, inner sep=0pt}]
  \tikzstyle{every node}=[font=\tiny]
  \def\N{80}
  \pgfmathparse{\N-1}
  \foreach \i [evaluate={\x=.9*\i/\N; \y=0.0}]in {0,...,\pgfmathresult}{ 
      \node[tsbluefill] (rv-\i) at (\x,\y) {};
  }
  \foreach \i [evaluate={\x=.9*\i/\N; \y=0.2}]in {0,...,\pgfmathresult}{ 
    \ifthenelse{\i = 4 \OR \i = 9 \OR \i = 14 \OR \i = 19 \OR \i = 24 \OR \i = 29 \OR \i = 34  \OR \i = 39 \OR \i = 44 \OR \i = 49 \OR \i = 54 \OR \i = 59 \OR \i = 64 \OR \i = 69 \OR \i = 74 \OR \i = 79}{
      \node[tsredfill] (rv5-\i) at (\x,\y) {};
    }{
      \node[tsgrayfill] (rv5-\i) at (\x,\y) {};
    }
  }
  
  \foreach \i [evaluate={\x=.9*\i/\N; \y=0.4}]in {0,...,\pgfmathresult}{ 
    \ifthenelse{\i = 19 \OR \i = 39 \OR \i = 59 \OR \i = 79}{
      \node[tsredfill] (rv20-\i) at (\x,\y) {};
    }{
      \node[tsgrayfill] (rv20-\i) at (\x,\y) {};
    }
  }
  
  \foreach \i [evaluate={\x=.9*\i/\N; \y=0.6}]in {0,...,\pgfmathresult}{ 
    \ifthenelse{\i = 39 \OR \i = 79}{
      \node[tsredfill] (rv40-\i) at (\x,\y) {};
    }{
      \node[tsgrayfill] (rv40-\i) at (\x,\y) {};
    }
  }
  
  \foreach \i [evaluate={\x=.9*\i/\N; \y=0.8}]in {0,...,\pgfmathresult}{ 
    \ifthenelse{\i = 79}{
      \node[tsredfill] (rv80-\i) at (\x,\y) {};
    }{
      \node[tsgrayfill] (rv80-\i) at (\x,\y) {};
    }
  }
  
  \foreach \i [evaluate={\x=.9*\i/\N; \y=1.0; \inext = int(\i + 1)}]in {0,...,\pgfmathresult}{ 
    \ifthenelse{\i = 79}{
      \node[tsgreenfill] (pred-\i) at (\x,\y) {};
    }{
      \node[tsgrayfill] (pred-\i) at (\x,\y) {};
    }
  }
  
  \foreach \iz in {4,9, 14, 19, 24, 29, 34, 39, 44, 49, 54, 59, 64, 69,74, 79}{
    \foreach \iy [evaluate={\ix=int(\iz+\iy);}] in {-4,-3,-2,-1,0}{
      \draw[thick] (rv-\ix.north) -- node[left] {} (rv5-\iz.south);
    }
  }
  \foreach \iz in {19, 39, 59, 79}{
    \foreach \iy [evaluate={\ix=int(\iz+5*\iy);}] in {-3,-2,-1,0}{
      \draw[thick] (rv5-\ix.north) -- node[left] {} (rv20-\iz.south);
    }
  }
  
  \foreach \iz in {39, 79}{
    \foreach \iy [evaluate={\ix=int(\iz+20*\iy);}] in {-1,0}{
      \draw[thick] (rv20-\ix.north) -- node[left] {} (rv40-\iz.south);
    }
  }
  
  \foreach \iz in {79}{
    \foreach \iy [evaluate={\ix=int(\iz+40*\iy);}] in {-1,0}{
      \draw[thick] (rv40-\ix.north) -- node[left] {} (rv80-\iz.south);
    }
  }
  \foreach \i in {79}{
    \draw[-,thick] (rv-\i.east) to [bend right=25] node[right] {} (pred-\i.east);
    \draw[-,thick] (rv5-\i.east) to [bend right=20] node[right] {} (pred-\i.east);
    \draw[-,thick] (rv20-\i.east) to [bend right=15] node[right] {} (pred-\i.east);
    \draw[-,thick] (rv40-\i.east) to [bend right=10] node[right] {} (pred-\i.east);
    \draw[-,thick] (rv80-\i.east) to [bend right=5] node[right] {} (pred-\i.east);
  }
\end{tikzpicture}}
\caption{Topologies of the NN models used in our experiments. The blue nodes in the bottom row denote the input time series while the green nodes in the top row denote the respective one-day-ahead predictions.}
\label{fig:harnets}
\end{figure}
The HARNet is based on a hierarchy of dilated causal convolutions and utilizes the recursive relationship defined in equation~\eqref{eq:rvavgrec}. For a multiplicatively nested sequence (cf.~\eqref{eq:multnest}) of $L$ aggregation periods $(j_1,\ldots,j_{L})$, the HARNet consists of $L$ convolutional layers, where the $l$-th layer is defined by a finite filter $\filt{l}$ and computes
\begin{equation}
\label{eq:layer}
f^{(l)}(x) = \begin{cases} \sigma\left(x \cconv \filt{l}\right),\;\;\; \filt{l}\in\mathbb{R}^{j_1} & \text{if } l = 1,\\\sigma\left(f^{(l-1)}(x)\cdconv{j_{l-1}} \filt{l}\right),\;\;\; \filt{l}\in\mathbb{R}^{j_{l}/j_{l-1}} & \text{else,} \end{cases}
\end{equation}
where $x\in\mathbb{R}^\mathbb{Z}$ is an input time series and $\sigma$ denotes the non-linear activation function. Here, we always use the rectified linear unit (ReLU) as an activation function, which is defined as
\begin{equation}
\label{eq:relu}
\sigma_{\mathrm{relu}}(x) = \begin{cases}x & \text{if } x > 0,\\
0 & \text{else}.
\end{cases}
\end{equation}
The complete function implemented by a HARNet is then given by
\begin{equation}
    \HARPhi{}(x) = \beta_0 + \sum_{l=1}^L\beta_l f^{(l)}(x),
\end{equation}
where the set $\theta$ of all trainable parameters consists of the weights $\beta_0, \ldots, \beta_L$ as well as the filters $\filt{1}, \ldots, \filt{L}$. Note, however, that in the case of $j_1 = 1$, which is true for all models considered in this work, we will omit the first convolutional layer and simply compute the identity instead. Then, the total number of model parameters can be computed as
\begin{equation}
|\theta| = 1 + L + \sum_{l=2}^L\frac{j_l}{j_{l-1}}.
\end{equation}
We will write $\HARPhi{20}$ and $\HARPhi{80}$ to denote the HARNet models that are based on the aggregation periods $(1, 5, 20)$ and $(1, 5, 20, 40, 80)$, respectively. The corresponding network topologies are shown in Figure~\ref{fig:harnets}. The models $\HARPhi{20}$ and $\HARPhi{80}$ thus contain $13$ and $19$ trainable parameters, respectively.
\subsection{Initialization of model parameters}
For a HARNet defined by a sequence of $L$ aggregation periods $(j_1, \ldots, j_L)$, we initialize the parameters $\beta_0, \ldots, \beta_L$ with the fitted parameters of the HAR model \eqref{eq:HAR}, which corresponds to the same sequence of aggregation periods. We consider three different techniques for estimating the baseline HAR model, namely OLS, WLS, and $\log$-OLS (cf. Section~\ref{sec:rvhar}).

The convolutional filter in the $l$-th layer is initialized with the average filter of the corresponding length, that is, 
\begin{equation}
    \filt{l} = \begin{cases}\avgf{j_1} & \text{if } l = 1,\\\avgf{j_l/j_{l-1}} & \text{else.} \end{cases}
\end{equation}

With this initialization and the ReLU as an activation function, the function $\HARPhi{}$ exactly replicates the predictions of the respective fitted HAR model, given that the input time series does not contain any negative values.

\subsection{Loss functions and optimization}
\label{sec:lossandopt}
\begin{table}[b]
\caption{Optimization hyperparameters. Choosing a batch size of $4$, and $5$ consecutive labels per training sample implies that a single batch contains labels for $20$ different one-day-ahead forecasts that are distributed across $4$ randomly sampled segments of the complete training time series. For a training time series that spans four years -- which roughly corresponds to \numprint{1000} labels -- these settings correspond to an approximate number of \numprint{200} epochs.\label{table:hyperparams}}
\centering
\begin{tabularx}{0.7\textwidth}{|X|c|}
\hline
Optimizer & ADAM\\ \cline{1-2}
Learning rate	& $1 \times 10^{-4}$\\ \cline{1-2}
Batch size & $4$\\ \cline{1-2}
Consecutive labels per training sample & $5$	\\ \cline{1-2}
Iterations & \numprint{10000}\\ \cline{1-2}
\hline
\end{tabularx}
\end{table}
Fitting the HAR model by OLS or WLS can be done via a closed formula and numerically only requires a single matrix inversion. However, even for linear problems such as the HAR model, this is only possible when considering the (weighted) mean squared error as the objective function. While gradient descent-based optimization is usually much more tedious and in most relevant situations not guaranteed to converge to a global minimum, it comes with the great advantage that it can be performed with arbitrary loss functions.

Here, we consider three different loss functions during optimization and for evaluating the forecast accuracy of trained models. These are the mean absolute error (MAE), the mean squared error (MSE) and the QLIKE error as proposed by \citet{robust_loss}. For an observation $y$ and a prediction $\widehat{y}$ produced by the model, these functions are defined by
\begin{align}
    \LMAE(\widehat{y}, y) &= \left|\widehat{y} - y\right|,\label{eq:lmae}\\
    \LMSE(\widehat{y}, y) &= \left(\widehat{y} - y\right)^2,\label{eq:lmse}\\
    \LQLIKE(\widehat{y}, y) &= \frac{y}{\widehat{y}} - \log\left(\frac{y}{\widehat{y}}\right) - 1.\label{eq:lqlike}
\end{align}
According to \citet{robust_loss}, $\LMAE$ and $\LQLIKE$ belong to a parametric family of robust and homogeneous loss functions for volatility forecasting, i.e., they are robust to noise in the volatility proxy and invariant to the choice of units of measurement.

For a given time series of daily RVs, HARNets are trained via mini-batch stochastic gradient descent. For each iteration of the gradient descent algorithm, a batch is constructed by randomly sampling a fixed number of training samples that are taken from different regions of the training time series. Each of these training samples is a segment of consecutive observations from the training time series whose length exceeds the length of the receptive field of the model by a fixed number of entries. Each of these additional entries defines an input-observation pair in the sense that the sample contains sufficient data for the model to compute the respective one-day-ahead prediction. This format is chosen because predictions for consecutive days can be computed efficiently during training due to the convolutional structure of the architecture. On the other hand, by considering batch sizes that are greater than one, it is also possible to combine samples from completely different parts of the training time series in a single batch.

In our numerical experiments, we apply \numprint{10000} iterations of the ADAM \citet{adam} optimizer with a fixed learning rate of $10^{-4}$. We use a batch size of $4$ and consider $5$ consecutive labels per training sample. That is, each input sample consists of a time series which exceeds the length of the receptive field of the model by $5$ entries, and a single batch thus contains a total of $20$ labels that come from $4$ different regions of the complete training time series.

In the subsequently presented results, we usually consider training time series that span four consecutive years, which corresponds to roughly \numprint{1000} one-day-ahead predictions that can be used as labels. This means that with \numprint{10000} iterations of stochastic gradient descent and $20$ labels per batch, our optimization algorithm runs for roughly \numprint{200} epochs. That is, each label in the complete training time series is used approximately \numprint{200} times. All hyperparameters are summarized in Table~\ref{table:hyperparams}.
\section{Results}
 \begin{figure}[p!]
    \centering
    \subfloat[]{\includegraphics[width=.85\linewidth]{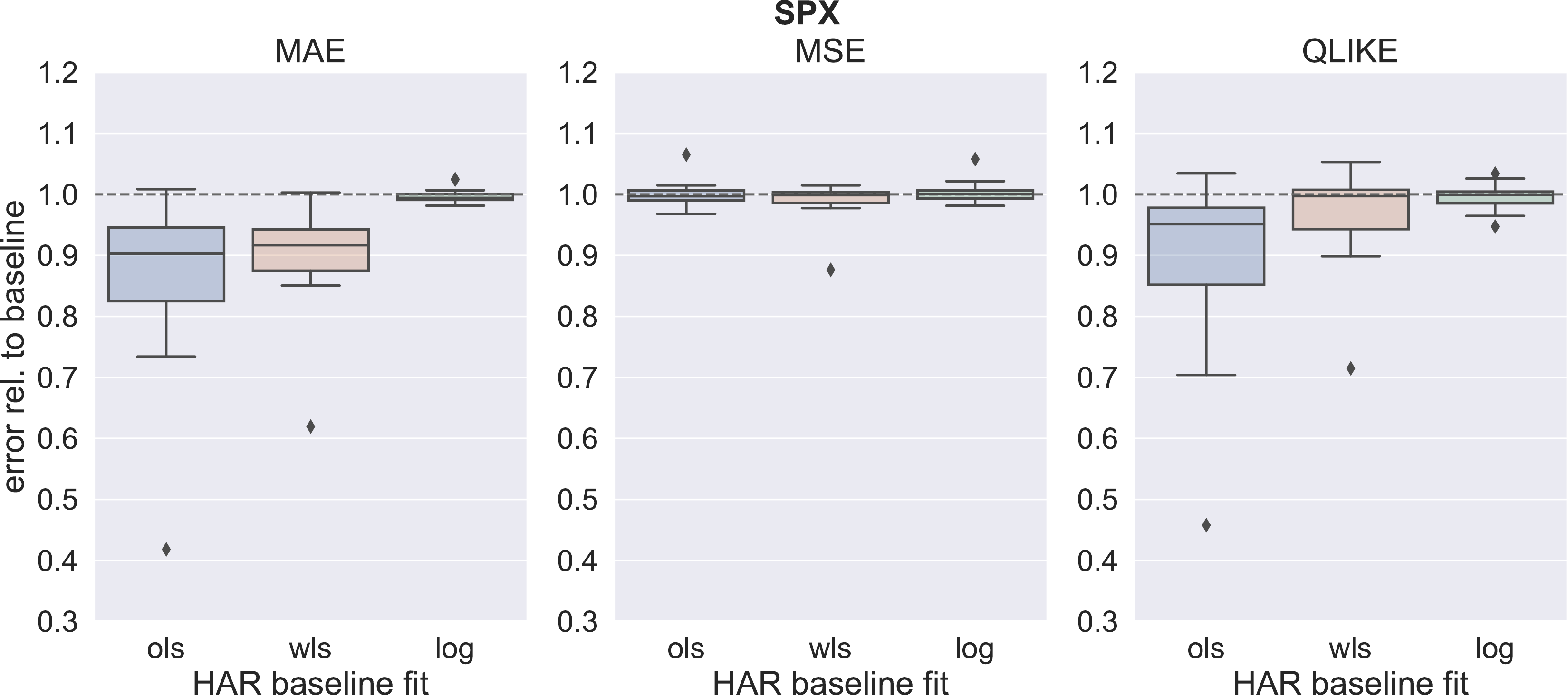}}\\[.5cm]
    \subfloat[]{\includegraphics[width=.85\linewidth]{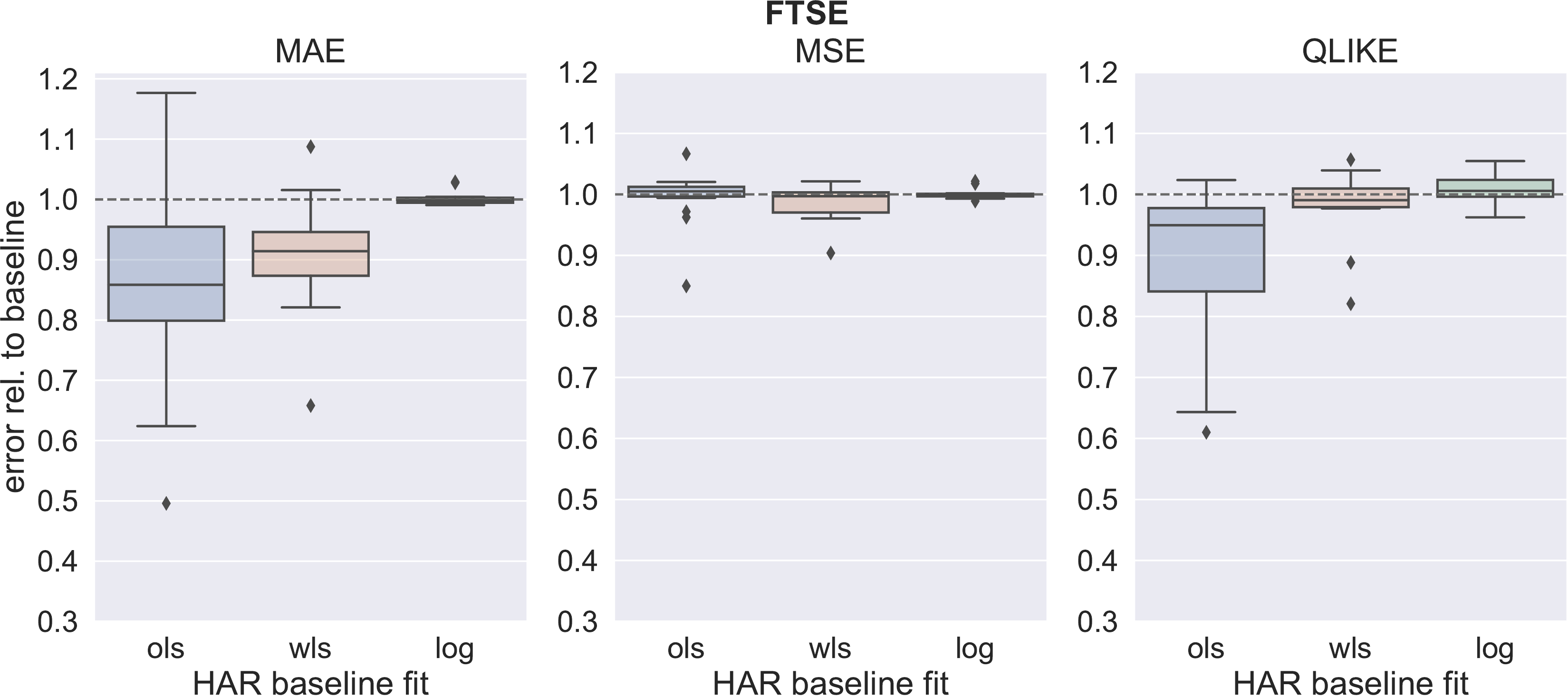}}\\[.5cm]
    \subfloat[]{\includegraphics[width=.85\linewidth]{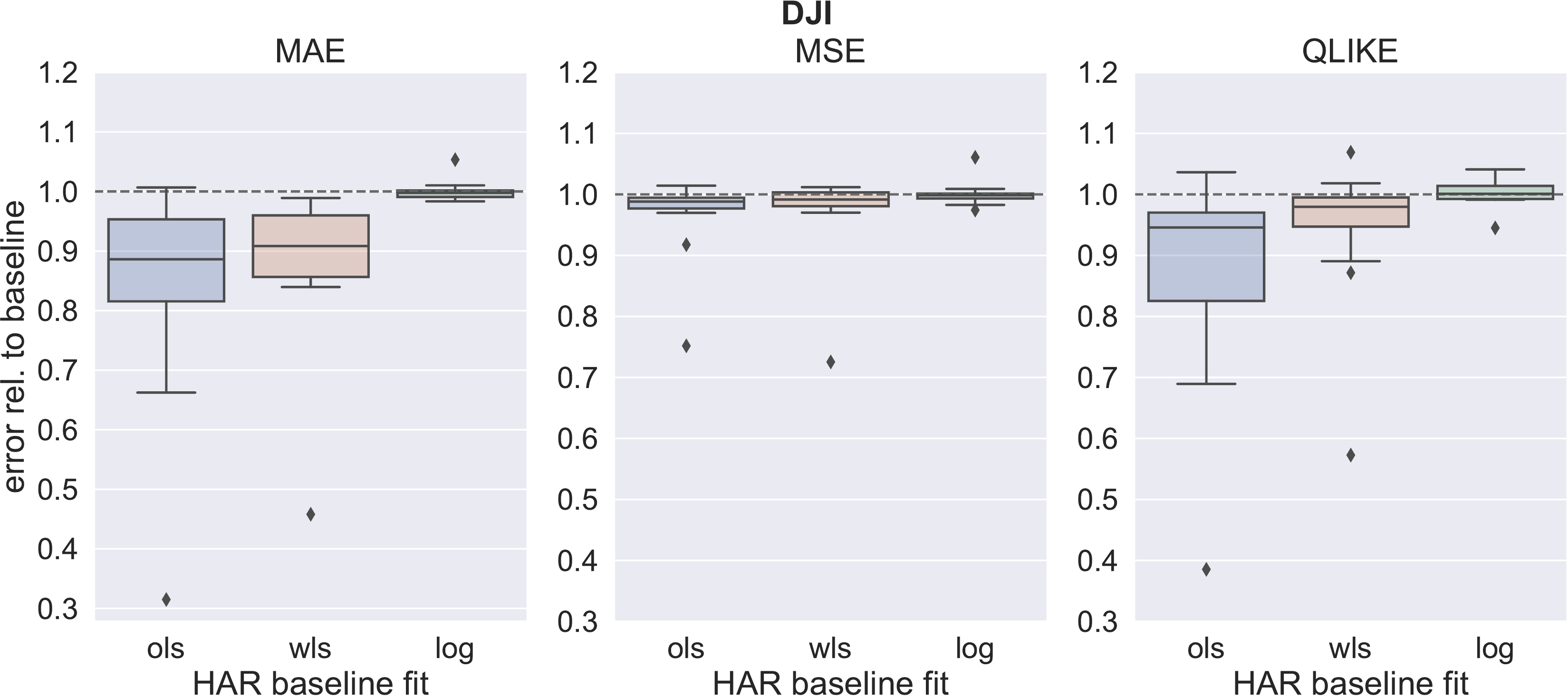}}
    \caption{Performance of the $\HARPhi{20}$ model with aggregation periods $(1, 5, 20)$ relative to the respective HAR baseline fit. Each boxplot depicts the distribution of the relative test errror for 15 training set/test set pairs taken from a time series of the respective index that ranges from 2002 to 2020.}
    \label{fig:base_20}
\end{figure}
\begin{figure}[p!]
    \centering
    \subfloat[]{\includegraphics[width=.85\linewidth]{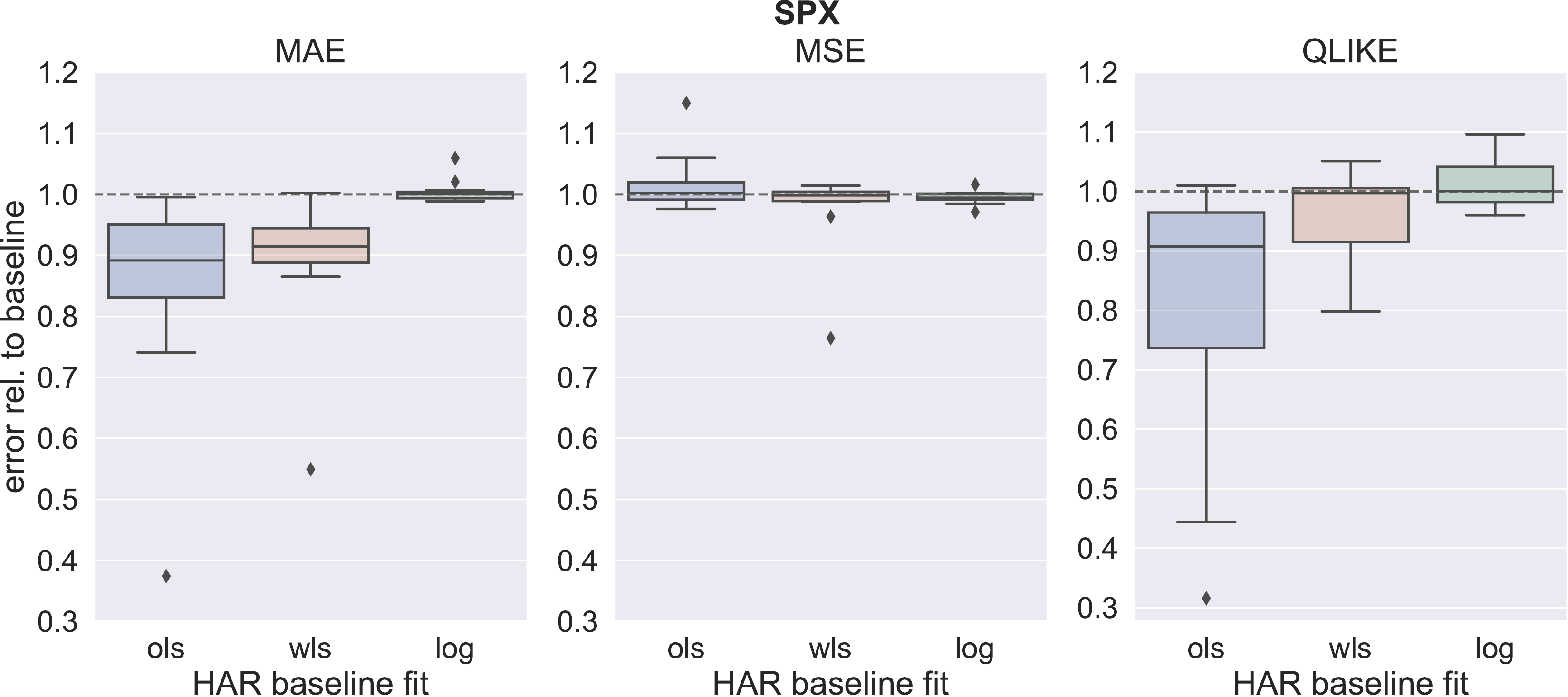}}\\[.5cm]
    \subfloat[]{\includegraphics[width=.85\linewidth]{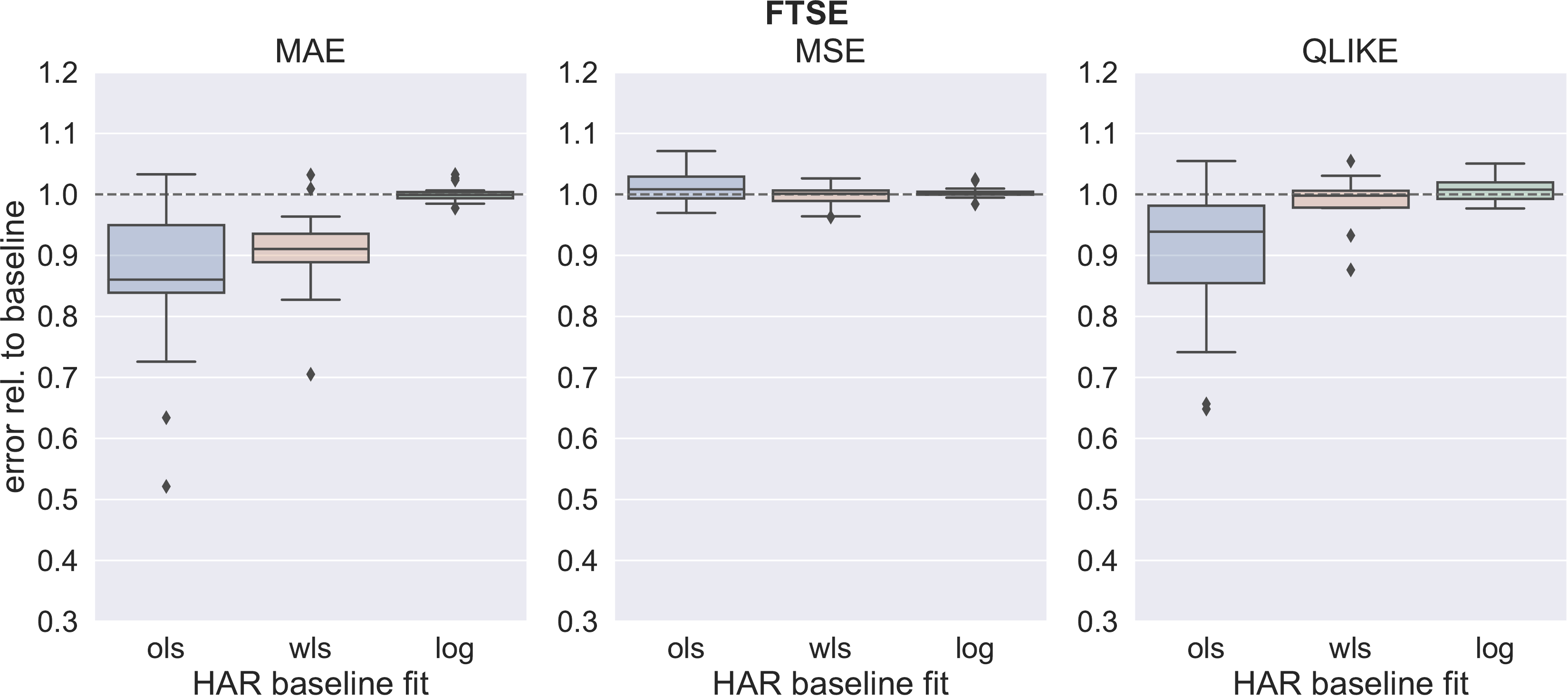}}\\[.5cm]
    \subfloat[]{\includegraphics[width=.85\linewidth]{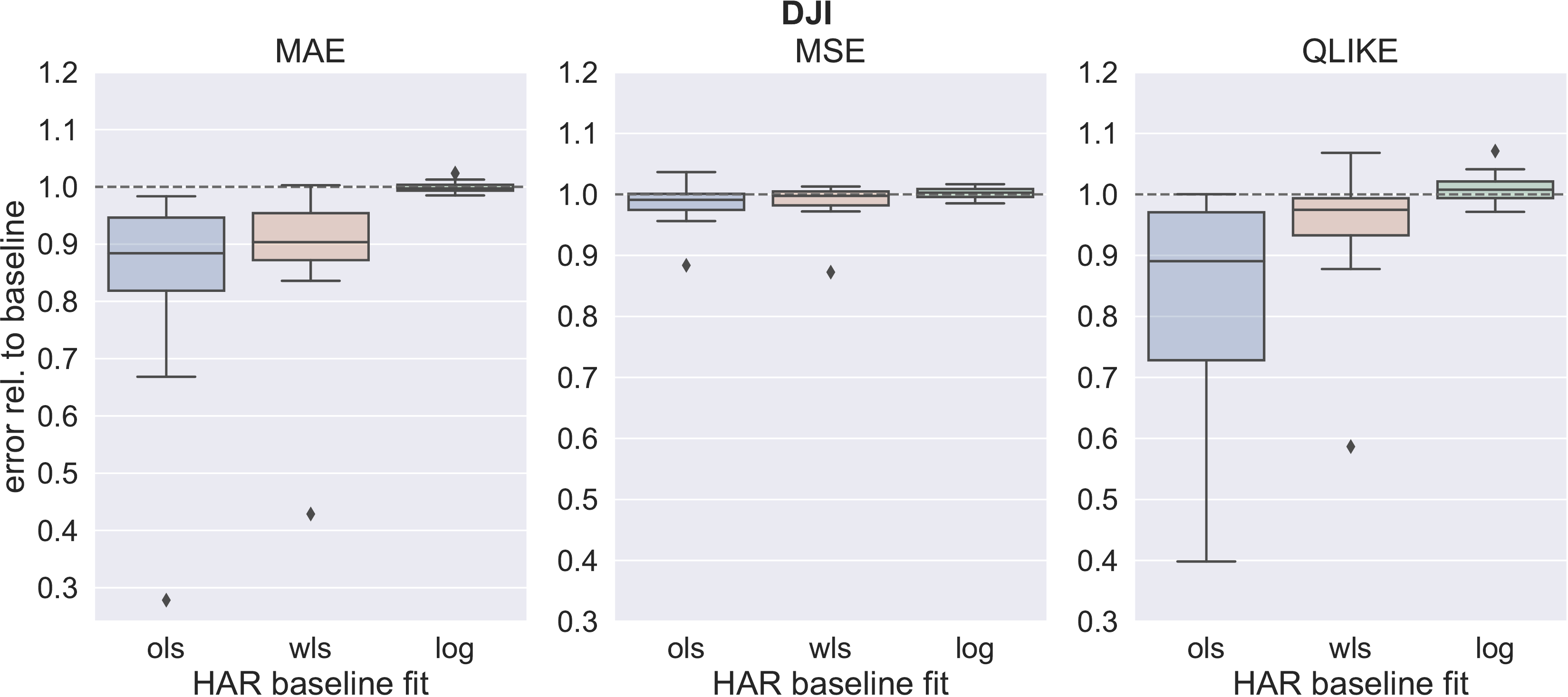}}
    \caption{Performance of the $\HARPhi{80}$ model with aggregation periods $(1, 5, 20, 40, 80)$ relative to the respective HAR baseline fit. Each boxplot depicts the distribution of the relative test errror for 15 training set/test set pairs taken from a time series of the respective index that ranges from 2002 to 2020.}
    \label{fig:base_80}
\end{figure}
\label{sec:results}
To evaluate the forecast accuracy of the proposed models and optimization techniques, we consider time series of daily RV for three different indexes, namely the S\&P500, the FTSE, and the DJI. The data is obtained from Oxford-Man Institutes realised library \parencite{realized_library}, from which we use the 5-min sub-sampled realized variance measure. Each of the three daily RV time series spans more than 19 years, ranging from 2001-08-29 to 2020-12-31.  We split each time series into 15 pairs of training and test sets, where each training set spans four consecutive years and the succeeding year is used as a test set. Throughout this section, we will compare different methods with respect to their performance across all 15 pairs of training and test sets for each of the three indexes.

We evaluate the HARNet models $\HARPhi{20}$ and $\HARPhi{80}$ by considering three different choices for the HAR baseline fit that is used for initializing the respective model parameters and three different loss functions. The considered loss functions are $\LMAE$, $\LMSE$, and $\LQLIKE$ (cf. Section~\ref{sec:lossandopt}). The three choices for the HAR baseline fit are OLS, WLS, and $\log$-OLS (cf. Section~\ref{sec:rvhar}). 
To ensure that predictions are strictly larger than zero, we clip the predicted value for both models at half of the smallest realized variance in the training set. All input RV time series are scaled linearly before being passed to the model such that $0$ is mapped to $0$ and $10^{-3}$ is mapped to $1$. When working with logarithmized time series, $-13$ is mapped to $0$ and $-2.5$ is mapped to $1$. Scaling input features to roughly cover the interval $[0,1]$ is common practice in many machine learning applications and often significantly improves gradient descent optimization.

Note that when using the $\log$-OLS fit of the HAR model as a baseline, the NN model is also optimized for the logarithmized time series. In this case, the model does not predict the realized variance, but the log-realized variance, which needs to be reconverted to the original scale of measurement. Here, we apply the bias correction proposed by \citet{log_corr}.

Figure~\ref{fig:base_20} depicts test errors relative to the test error of the respective HAR baseline fit for the $\HARPhi{20}$ model after optimization with three different loss functions. In each case, the error on the test set is computed with the same loss function, which was already used during optimization. Note that before optimization, $\HARPhi{20}$ and the HAR baseline model yield identical predictions. Our results therefore provide a detailed quantitative analysis as to how the performance of the model changes from the HAR baseline during optimization. Analogous results for the $\HARPhi{80}$ model are shown in Figure~\ref{fig:base_80}.

For both models $\HARPhi{20}$ and $\HARPhi{80}$, we observe that the largest relative improvement can be achieved when using $\LMAE$ as a loss function and the standard OLS approach for fitting the baseline HAR model. In the case of $\HARPhi{20}$, across all indexes, we achieve an average reduction of the median test MAE of about 11.74\%. Substantial improvements in terms of the MAE can also be observed when choosing the WLS fit as a starting point. However, when considering logarithmized RV time series, the relative improvement in accuracy is greatly reduced. Using $\LQLIKE$ as a loss functions yields slightly reduced but still considerable improvements relative to the HAR baseline fit, in particular when considering the standard OLS approach. In the case of $\LMSE$, however, the test error of the NN model relative to the baseline HAR fit remains almost the same after optimization.

Based on the results presented in Figures \ref{fig:base_20} and \ref{fig:base_80}, we believe that the following observations are of significant importance and merit further analysis:
\begin{enumerate}
    \item While \citet{poon2020machine} do not recommend using the QLIKE error as an objective function during optimization, we found that this approach can yield consistent and substantial improvements. Can these diverging experimental results be explained by our approach of initializing the parameters of the HARNet model?
    \item The QLIKE error has been described as a superior metric for ranking different volatility forecasting models \parencite{patton2009optimal}. Does applying $\LQLIKE$ as a loss function also yield improvements with respect to other metrics such as the MAE or the MSE and can thus be generally recommended in this setting?
    \item Our method yields substantial improvements relative to the baseline HAR model when considering the standard OLS fit but often little to no improvements when using WLS or $\log$-OLS. This raises the question whether HARNets that used an OLS fit as a starting point also outperform HAR models that were fitted with an WLS or $\log$-OLS approach.
    \item On average, the optimized $\HARPhi{20}$ model matches the performance on the test data of the respective baseline HAR fit and often clearly outperforms it. That is, it often generalizes better despite having significantly more model parameters. In particular, $\HARPhi{20}$ learns weighted filters for the weekly and monthly aggregation periods, whereas the original HAR model only considers fixed average filters. Can an analysis of the learnt filter weights provide further insights as to which aggregates of past data are of significant importance when forecasting RV? 
\end{enumerate}
In the following subsections, we will perform a detailed analysis of the questions raised above.
\subsection{Initialization with a HAR baseline model stabilizes optimization with the QLIKE loss}
\begin{figure}[th!]
    \centering
    \subfloat[\label{fig:rand_init}Random initialization of weights.]{\includegraphics[width=.85\linewidth]{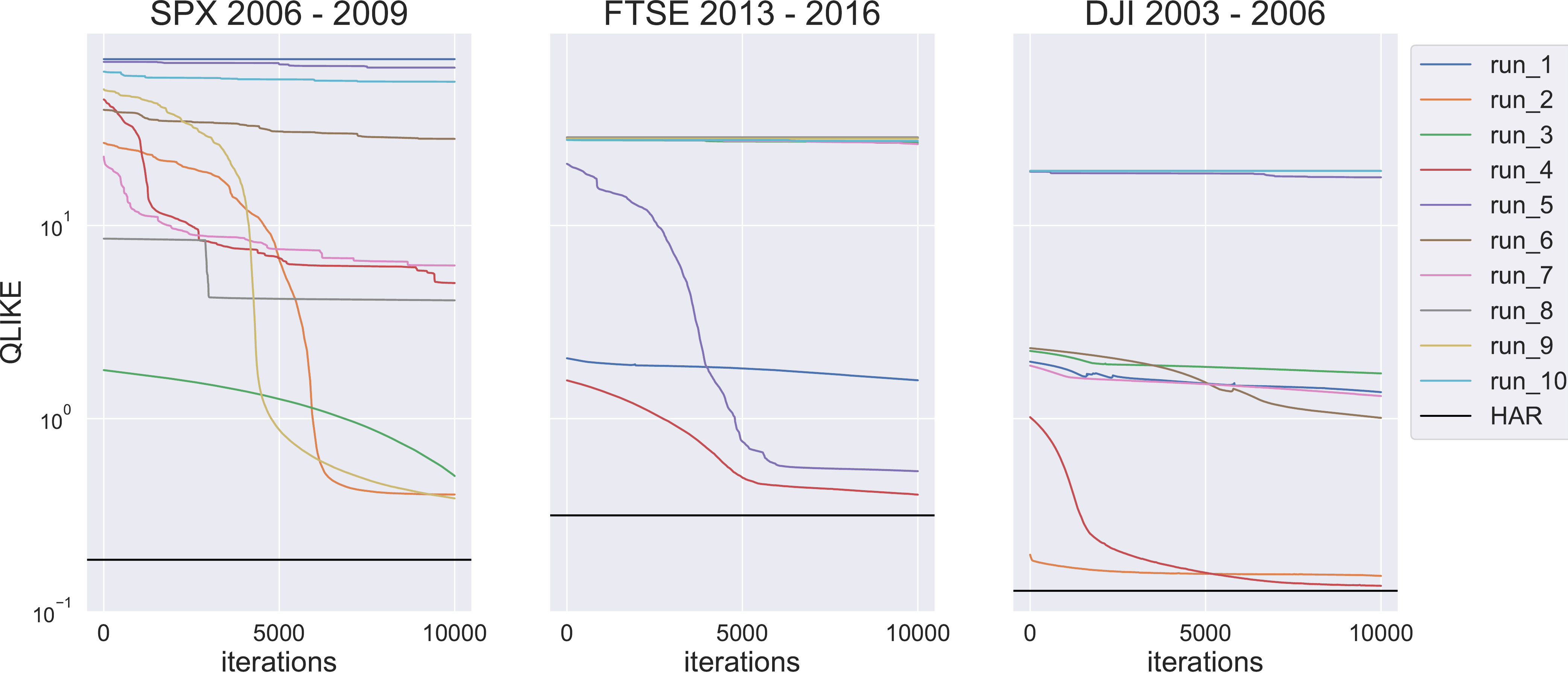}}\\[.5cm]
    \subfloat[\label{fig:har_init}Proposed initialization with the HAR baseline model.]{\includegraphics[width=.85\linewidth]{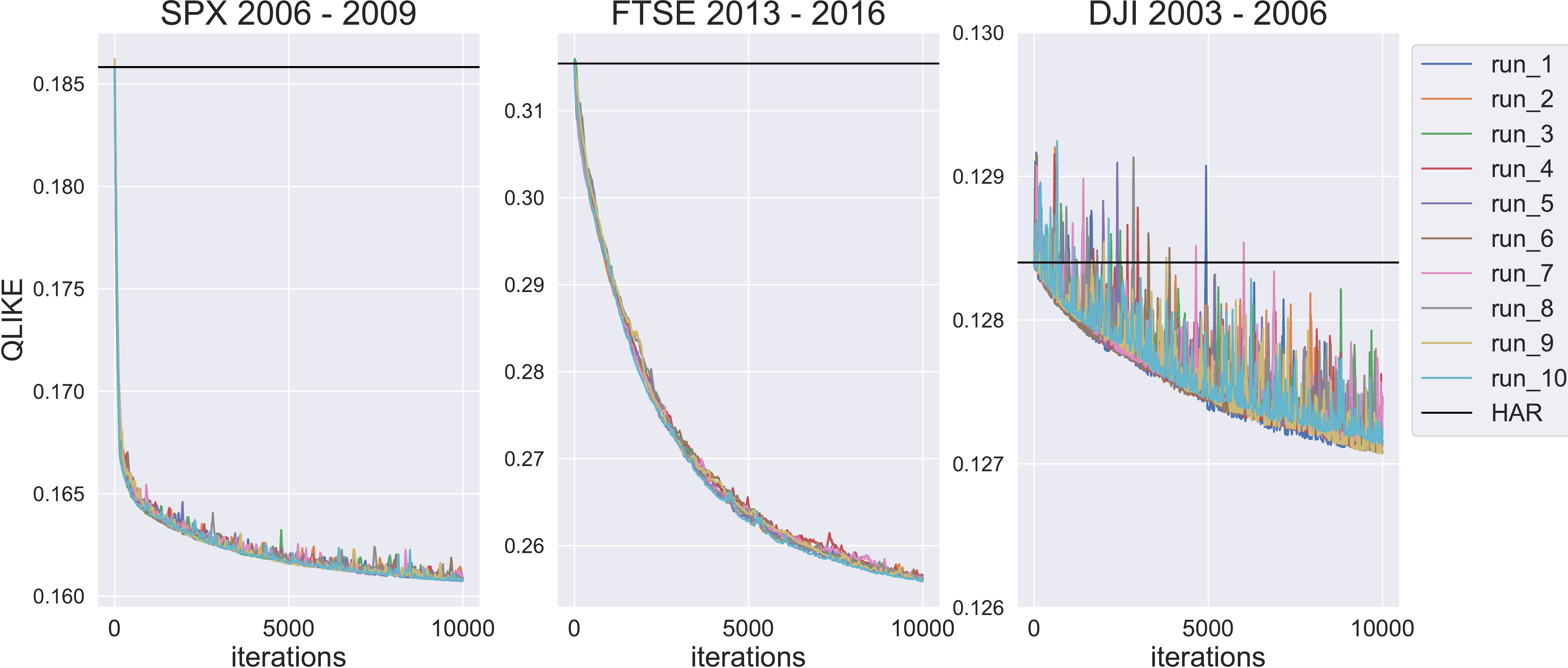}}
   \\[.5cm]
\subfloat[\label{tab:harvsrandqlike}Summary of QLIKE training errors after \numprint{10000} iterations.]{\resizebox{\textwidth}{!}{
\begin{tabular}{|r|lll|lll|lll|}
\hline
&\multicolumn{3}{c|}{SPX 2006 - 2009}                         & \multicolumn{3}{c|}{FTSE 2013 - 2016} & \multicolumn{3}{c|}{DJI 2003 - 2006} \\ 
QLIKE &
  \multicolumn{1}{c}{median} &
  \multicolumn{1}{c}{mean} &
  \multicolumn{1}{c|}{std} &
  \multicolumn{1}{c}{median} &
  \multicolumn{1}{c}{mean} &
  \multicolumn{1}{c|}{std} &
  \multicolumn{1}{c}{median} &
  \multicolumn{1}{c}{mean} &
  \multicolumn{1}{c|}{std} \\  \hline
Random init                    & 5.62038 & 23.84032 & 28.00004 & 27.08056    & 19.66301   & 12.34607   & 1.54270    & 8.09141    & 8.76879   \\
HAR baseline init & 0.16084 & 0.16083  & 0.00005  & 0.25620     & 0.25621    & 0.00021    & 0.12720    & 0.12722    & 0.00013   \\ \hline
\end{tabular}}}
\caption{Initialization with a baseline HAR model stabilizes the optimization of a HARNet when using the QLIKE error as a loss function. (a) Optimization curves for ten different optimization runs of the $\HARPhi{20}$ model with randomly initialized weights. (b) Optimization curves for ten different optimization runs of the $\HARPhi{20}$ model with the proposed initialization with a HAR baseline model. (c) QLIKE training errors after \numprint{10000} iterations of the ADAM algorithm.\label{fig:harvsrandqlike}}
\end{figure}
\citet{poon2020machine} report an inferior performance of machine learning models for RV forecasting that were optimized with a QLIKE loss as compared to an MSE loss. They argue that this could be due to the insensitivity of the QLIKE error in cases where the true RV is very high. In practice, such a flat loss function would make it very difficult for the optimizer to overcome large training errors on volatility jump days during optimization.

We suspect that this issue can at least partially be overcome in our setting due to the proposed initialization approach. After initialization, the HARNet yields identical predictions as the baseline HAR model. Thus, it can produce extreme prediction errors only to the same extent as produced by the baseline model. However, when considering a random initialization of the model weights, this is not the case and extreme prediction errors are very common during early optimization.

To test this hypothesis experimentally, we compare optimization curves for the model $\HARPhi{20}$ when using random initialization techniques and when using an OLS fit of the HAR model as a baseline. For random initialization, we consider the Glorot uniform initializer \parencite{glorot2010understanding}, which is one of the most common initialization methods for NN weights. The Glorot uniform initializer draws samples from a uniform distribution over an interval whose limits depend on the number of input and output units in the weight tensor of a layer.

Figure~\ref{fig:harvsrandqlike} contains the results of an experiment where we first randomly selected a time span of four years as a training set for each of the three indexes, and then performed 10 independent optimizations with both initialization methods, respectively. Our results clearly show that the QLIKE training error highly depends on the initialization when considering random initialization. As shown in Figure~\ref{fig:rand_init}, random initialization causes a high variation in the final training error after \numprint{10000} iterations of the ADAM algorithm, while all of the 10 optimization runs yield training errors that are greater than the training error of the fitted HAR model with the same aggregation periods. When considering initialization with a baseline HAR model, on the other hand, all of the 10 optimization curves follow basically the same path and improve upon the QLIKE training error of the baseline model (cf. Figure~\ref{fig:har_init}).
\subsection{QLIKE optimization also generalizes with respect to the MAE and MSE test metrics}
\label{sec:qlikegen}
\begin{figure}[th!]
    \centering
    \subfloat[HARNet with aggregation periods $(1, 5, 20)$.]{\includegraphics[width=.85\linewidth]{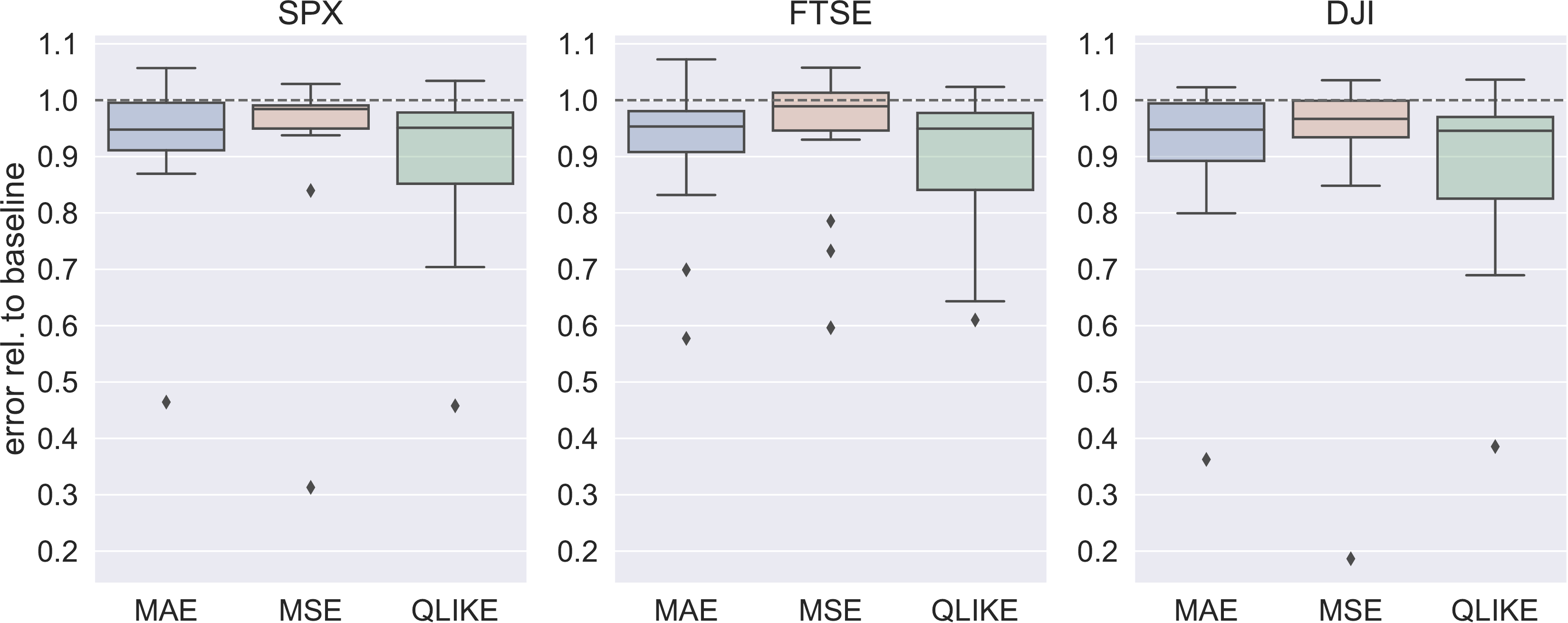}}\\[0.5cm]
    \subfloat[HARNet with aggregation periods $(1, 5, 20, 40, 80)$.]{\includegraphics[width=.85\linewidth]{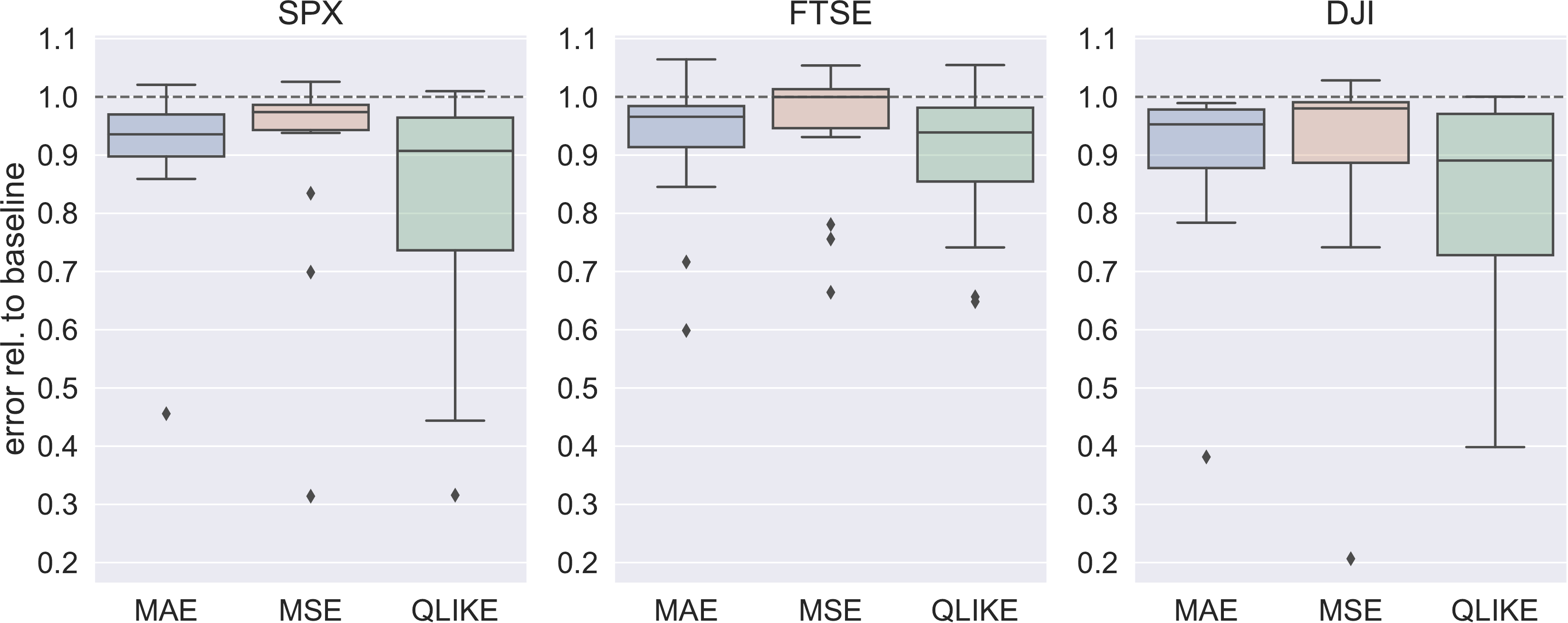}}
    \caption{Performance of the $\HARPhi{20}$ and $\HARPhi{80}$ models after optimization with the QLIKE loss function in terms of the MAE, MSE, and the QLIKE test error relative to the HAR baseline fit. The baseline fit was obtained by OLS. Each boxplot depicts the distribution of the relative test errror for 15 training set/test set pairs taken from a time series of the respective index that ranges from 2002 to 2020.}
    \label{fig:qlike_optim}
\end{figure}
The QLIKE error is often considered to be the preferred evaluation metric in the context of RV forecasting. In our initial results, we found that when using the QLIKE error as a loss function, a HARNet can substantially improve upon the QLIKE test error of a HAR baseline model which was fitted by OLS (cf. Figures~\ref{fig:base_20}~and~\ref{fig:base_80}). This raises the question whether optimizing with a QLIKE loss function can also improve other test metrics such as the MAE or the MSE.

Our results compiled in Figure~\ref{fig:qlike_optim} indeed show that HARNets that were optimized with a QLIKE loss not only yield better test errors with respect to the QLIKE metric than the baseline HAR model but also with respect to the MAE and the MSE. While the improvements in terms of the MAE are smaller than when directly optimizing with an MAE loss, we found that this approach leads to consistent albeit small improvements in terms of the MSE test error. This is particularly surprising as in our initial results, we found that using the MSE as a loss function usually fails at improving the MSE test error of the baseline model. 
\subsection{HARNets that are initialized with an OLS fit of the HAR baseline model outperform WLS and log-OLS fits of the baseline model}
\label{sec:olsbest}
\begin{figure}[p]
    \centering
    \subfloat[HARNet with aggregation periods $(1, 5, 20)$.]{\includegraphics[width=.8\linewidth]{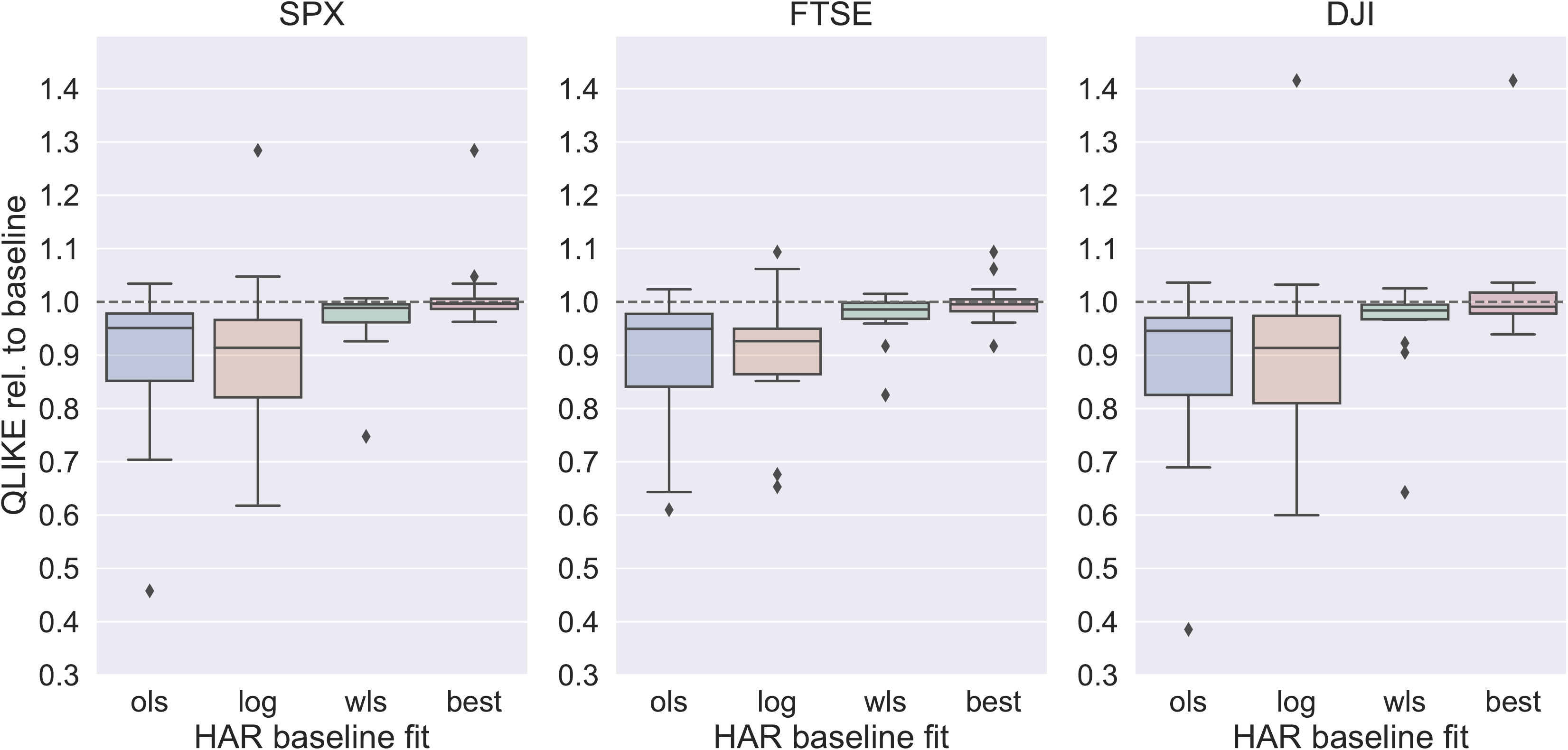}}\\[.5cm]
    \subfloat[HARNet with aggregation periods $(1, 5, 20, 40, 80)$.]{\includegraphics[width=.8\linewidth]{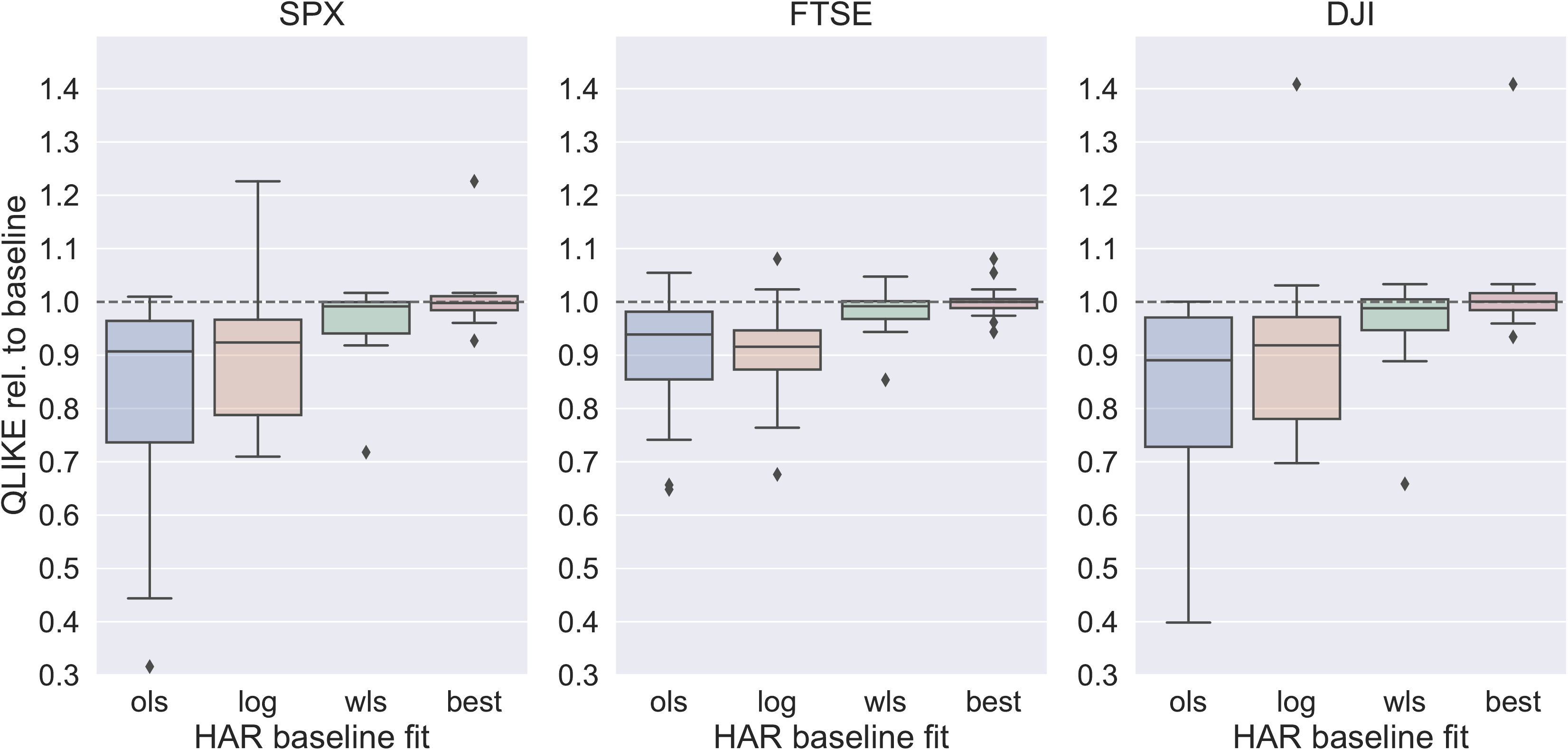}}\\[.5cm]
    \subfloat[\label{fig:loss_opt}A HARNet model initialized with the OLS fit of the baseline model outperforms the OLS, WLS, and $\log$-OLS fit of the baseline model on the test set after optimization ]{\includegraphics[width=.6\linewidth]{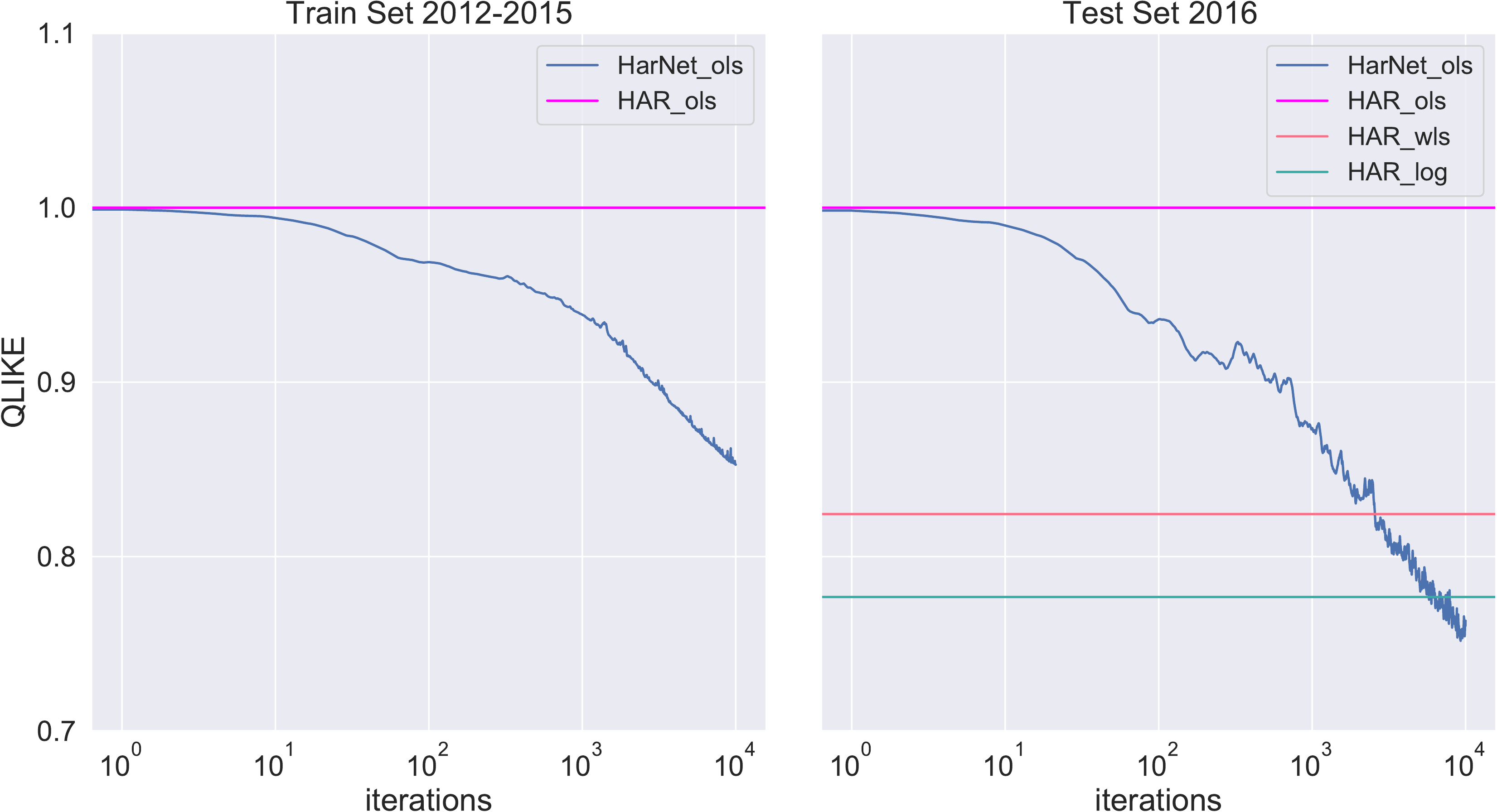}}
    \caption{Performance of the $\HARPhi{20}$ and $\HARPhi{80}$ models relative to different fits of the HAR baseline model. All HARNets were initialized with the OLS fit of the baseline model and optimized with the QLIKE loss function. The method \emph{best} means that the decision whether to use the OLS, WLS, or $\log$-OLS approach for fitting the baseline model used in the comparison with the respective HARNet was made \emph{after} knowing the test errors. Each boxplot depicts the distribution of the relative test errror for 15 training set/test set pairs taken from a time series of the respective index that ranges from the year 2002 until 2020.}
    \label{fig:qlike_optim_vs_har_fits}
\end{figure}
Our main results presented in Figures~\ref{fig:base_20}~and~\ref{fig:base_80} show that HARNets can substantially improve upon the forecasting accuracy when using an OLS fit of the HAR baseline model. However, this is in general not true when fitting the baseline model by WLS, while using the log-OLS fit of the baseline as a starting point for the HARNet usually yields no significant changes in the accuracy on the test set whatsoever. 

The results presented in Figure~\ref{fig:qlike_optim_vs_har_fits} show that HARNets that were initialized with the OLS fit of the baseline model also consistently outperform the respective WLS and $\log$-OLS fits of the baseline model in terms of the QLIKE error. We even found that HARNets that were initialized with the OLS fit of the baseline model consistently match the performance of the HAR baseline model which was obtained by the \emph{best} method for fitting its parameters, i.e., for which the decision whether to use OLS, WLS, or $\log$-OLS was made after knowing the respective test errors.

In the $\log$-OLS approach, the training time series is logarithmized before solving the linear system. This transformation introduces non-linearity to the forecasting model and implicitly reduces the relative importance of days with a very high RV in the OLS fit. When using WLS, on the other hand, one explicitly defines weights, which usually also reduce the importance of days of RV jumps. Both approaches are explicitly defined schemes that aim to overcome the limitations of simple linear models, such as the HAR model, without adding any complexity to the model itself. We believe that the results presented in this section suggest that, due to the significantly increased expressiveness, the HARNet model is capable of learning similarly effective relationships directly from the data. Explicitly incorporating the WLS or $\log$-OLS fit of the HAR baseline model, however, seems to either cause the HARNet model to get stuck in unfavorable local minima or increase the risk of overfitting.

In summary, our findings suggest that when used for forecasting of daily RV, HARNets are best optimized with a QLIKE loss function and an initialization scheme which uses the OLS fit of the HAR baseline model as a starting point.
\subsection{Interpretability of HARNet filter weights}
\label{sec:interpret}
\begin{figure}[th!]
    \centering
    \includegraphics[width=.85\linewidth]{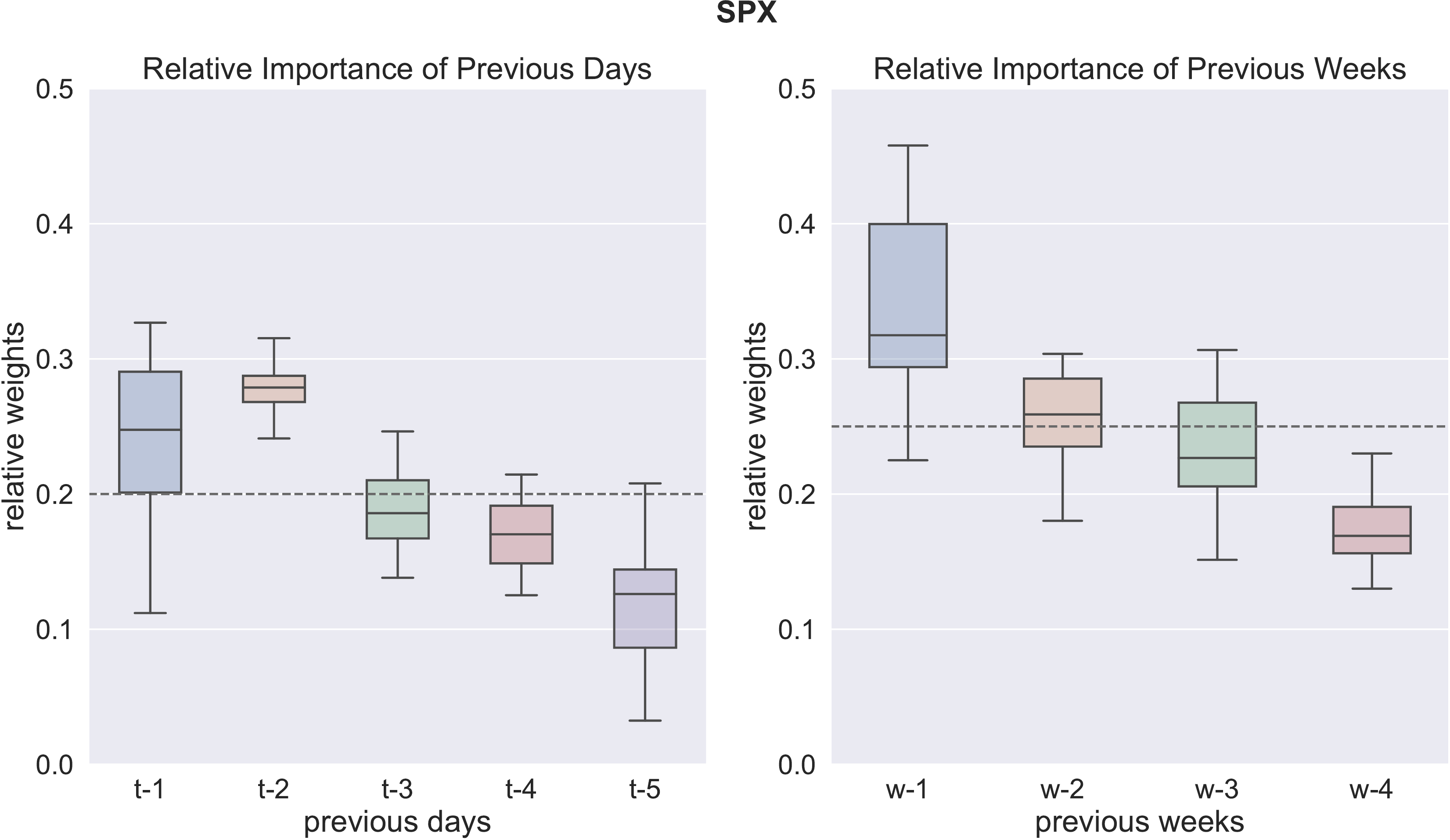}
    \caption{Distribution of normalized filter weights in the layers of a $\HARPhi{20}$ model which was initialized with the OLS fit of the HAR baseline model and optimized with a QLIKE loss function for the 15 training sets obtained from the SPX time series.}
    \label{fig:weights_change}
\end{figure}
Due to the close relationship to the original HAR model, HARNets allow for a high degree of interpretability when forecasting RV, setting them apart from generic machine learning models which are often treated as black boxes. A particularly interesting property of a HARNet is that the time series corresponding to a certain aggregation period is computed via weighted averages rather than normal averages, as it is the case in the original HAR model. The filter kernels which define the respective weights are learnt during optimization and can thus inform us as to which of the previous days or weeks are in fact significant for RV forecasting within a given training time series.

Figure~ \ref{fig:weights_change} depicts the distribution of the weights in the two convolutional layers of a $\HARPhi{20}$ model which was initialized with the OLS fit of the HAR baseline model and optimized with a QLIKE loss function for the 15 training sets obtained from the SPX time series. With respect to the first layer, it can be observed that the HARNet strongly focuses on the RV at time points $t-1$ and $t-2$, while the importance of the other days within the past week gradually diminishes. In particular, the RV at the day before yesterday played a significant role in all fits of the model. A possible explanation for this is that single RV jumps can only be distinguished from periods of generally high RV when at least considering the values at $t-1$ and $t-2$ in the time series. In the second layer, we can observe an almost linearly diminishing importance of the weeks in the past month.
\section{Additional regressors}
\begin{figure}[th!]
    \centering
    \subfloat[Extended HARNet with aggregation periods $(1, 5, 20)$.]{\includegraphics[width=.8\linewidth]{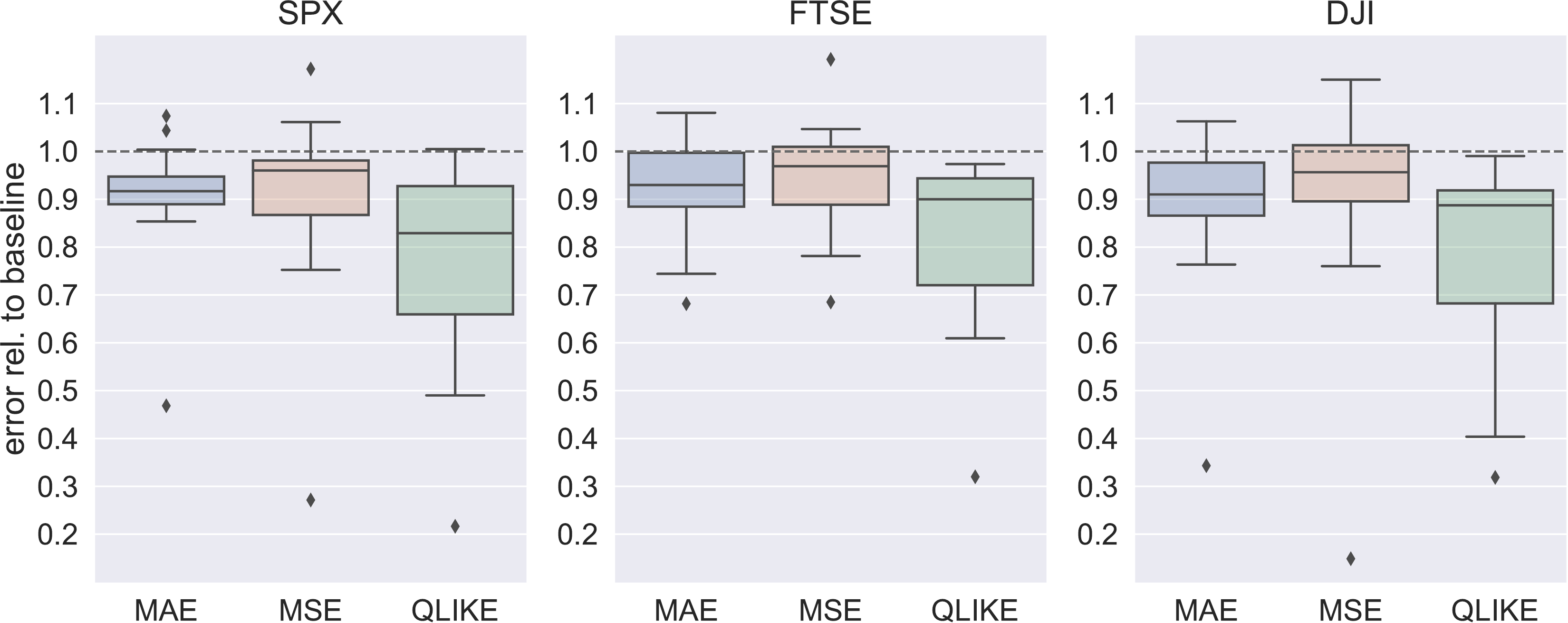}}\\[0.5cm]
    \subfloat[Extended HARNet with aggregation periods $(1, 5, 20, 40, 80)$.]{\includegraphics[width=.8\linewidth]{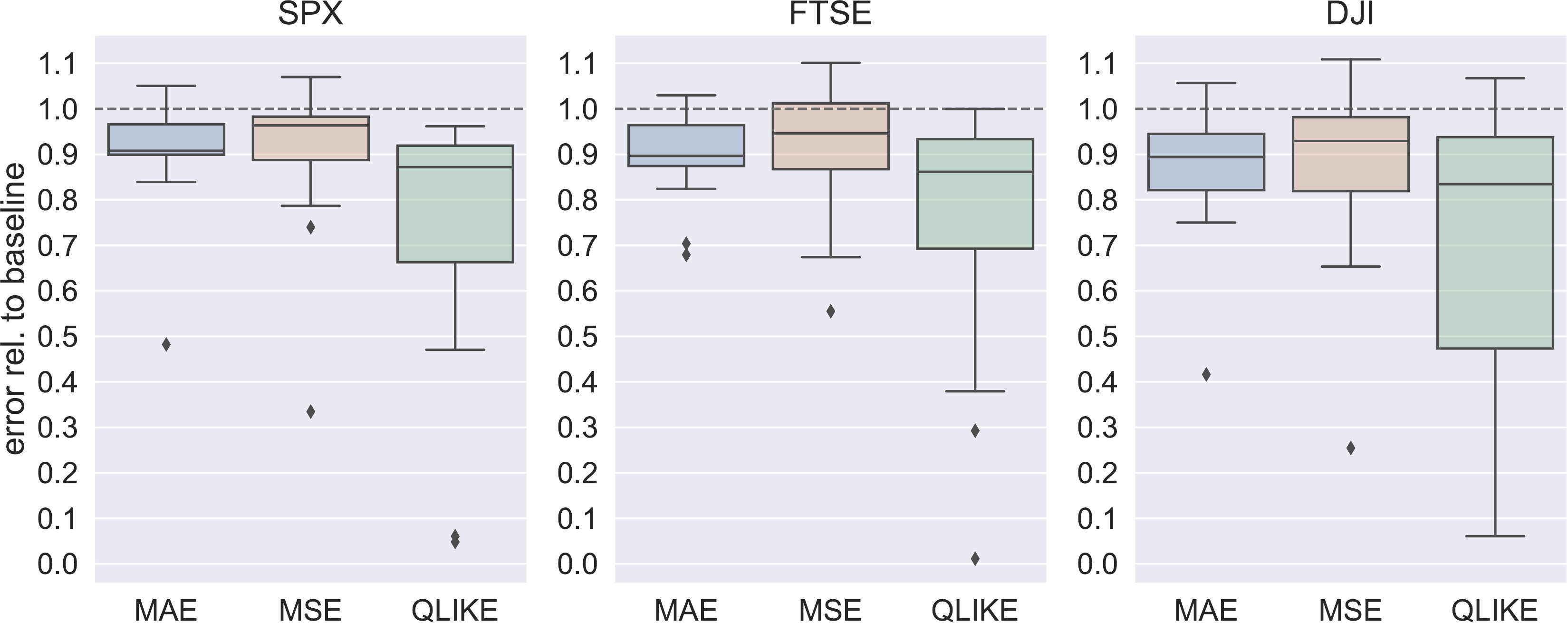}}
    \caption{Performance of the extended $\HARPhi{20}$ and $\HARPhi{80}$ models after optimization with the QLIKE loss function in terms of the MAE, MSE, and the QLIKE test error relative to the HAR baseline fit. The baseline fit was obtained by OLS. The input time series additionally contained values for the daily realized semivariance and the signed jump variation. Each boxplot depicts the distribution of the relative test errror for 15 training set/test set pairs taken from a time series of the respective index that ranges from the year 2002 until 2020.}
    \label{fig:additional_base}
\end{figure}
\begin{figure}[th!]
    \centering
    \subfloat[Extended HARNet with aggregation periods $(1, 5, 20)$.]{\includegraphics[width=.8\linewidth]{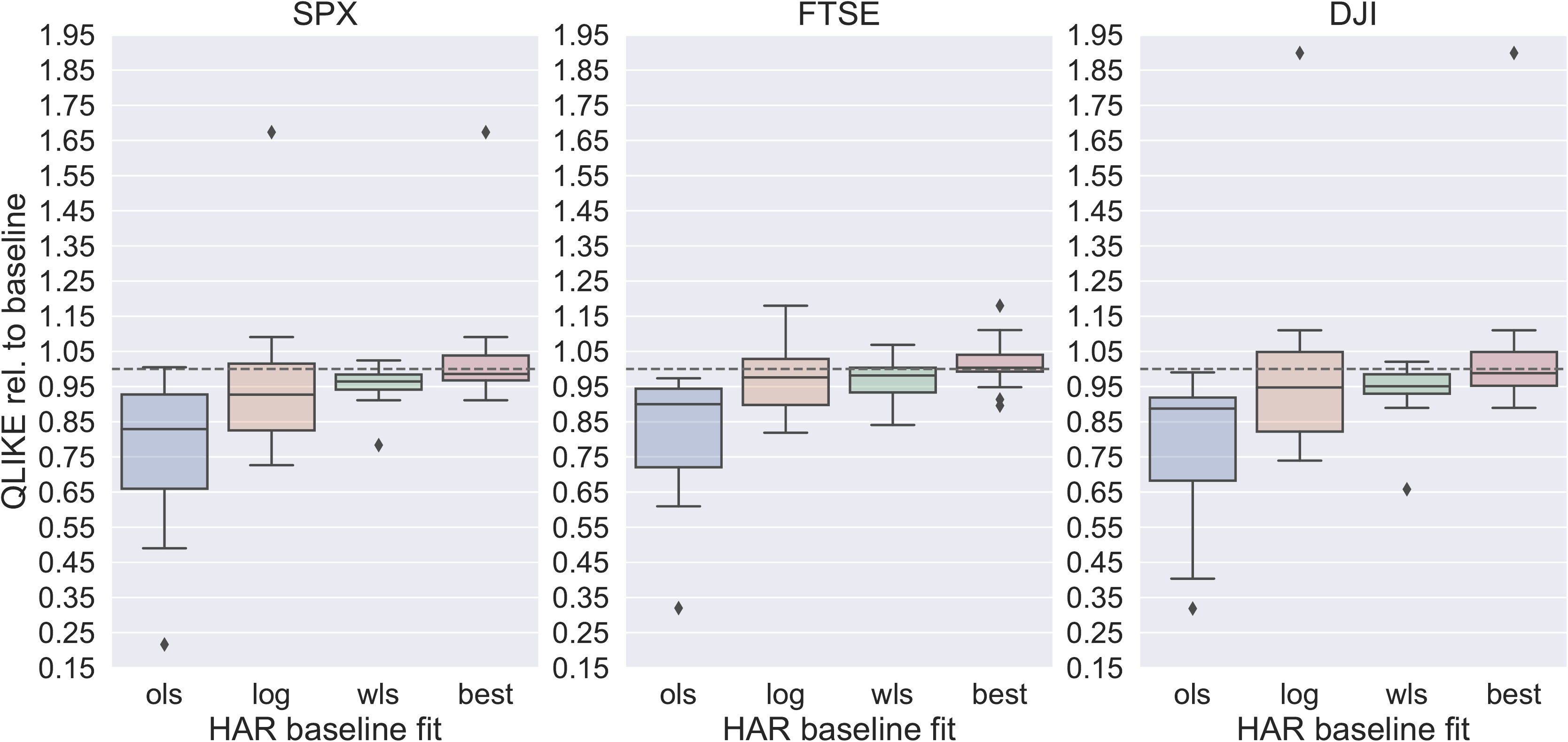}}\\[0.5cm]
    \subfloat[Extended HARNet with aggregation periods $(1, 5, 20, 40, 80)$.]{\includegraphics[width=.8\linewidth]{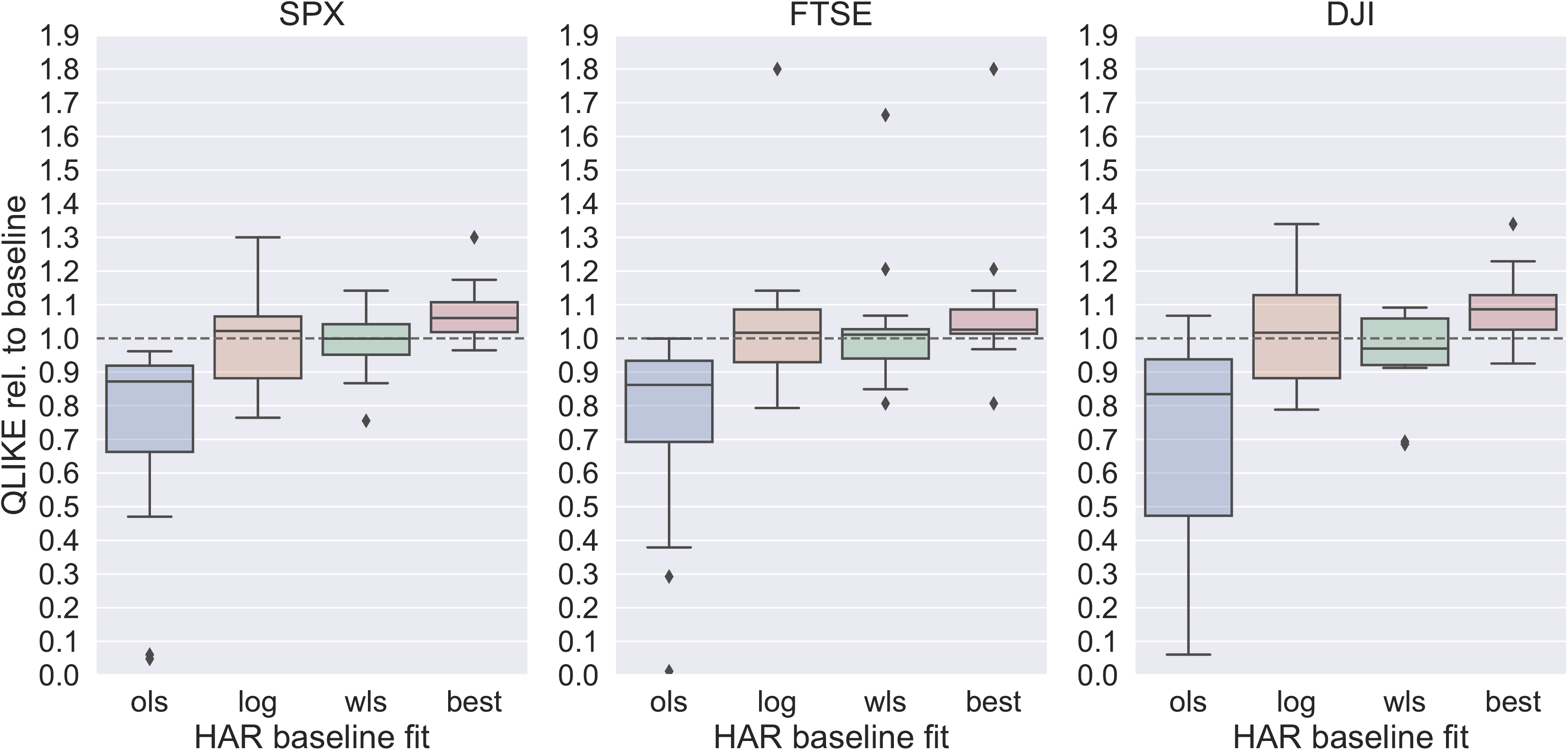}}
    \caption{Performance of the extended $\HARPhi{20}$ and $\HARPhi{80}$ models relative to different fits of the HAR baseline model. All HARNets were initialized by OLS estimation of the baseline model and optimized with the QLIKE loss function. The method \emph{best} means that the decision whether to use the OLS, WLS, or $\log$-OLS approach for fitting the baseline model used in the comparison with the respective HARNet was made \emph{after} knowing the test errors. The input time series additionally contained values for the daily realized semivariance and the signed jump variation. Each boxplot depicts the distribution of the relative test errror for 15 training set/test set pairs taken from a time series of the respective index that ranges from the year 2002 until 2020..}
    \label{fig:additional_vs}
\end{figure}
\label{sec:semivar}
Incorporating additional information contained in the time series of high-frequency returns can significantly improve the accuracy of RV forecasting models. Motivated by \citet{patton_good_2015} we consider the negative realized semivariance and the signed jump variation as additional inputs to our models. The ability of neural networks to identify nonlinear relations in the data might prove beneficial in a setting where this type of information is available. In line with \citet{semivariance}, the realized semivariance decomposes the realized variance into  components associated with
positive and negative returns, respectively. Analogously to $RV_t$ (cf.~equation~\eqref{eq:rv}), these components are defined as
\begin{align}
    \RS_{t}^+ &= \sum_{n=0}^{N-1} r_{t,n}^2 \mathds{1}_{r_{t,n} >0},\\
    \RS_{t}^- &= \sum_{n=0}^{N-1} r_{t,n}^2 \mathds{1}_{r_{t,n} < 0},
\end{align}
where $r_{t,n}$ are the intraday returns and $\mathds{1}$ denotes the indicator function. We denote $\RS_{t}^{+(j)}$ and $\RS_{t}^{-(j)}$ as the respective $j$-day averages. The signed jump variation is defined as the difference between $\RS_{t}^+$ and $\RS_{t}^-$, that is,
\begin{equation}
    \Delta J_t^2 =  \RS_t^+ - \RS_t^-.
\end{equation}
Accordingly, we consider the following linear HAR-type baseline model:
\begin{align}
\begin{split}
\label{eq:HARRSJ}
    RV_t = \beta_0 &+ \beta_1 \RV_{t-1} + \beta_2 \RV^{(5)}_{t-1} + \beta_3 \RV^{(20)}_{t-1} \\
    &+ \beta_4 \RS_{t-1}^- + \beta_5 \RS_{t-1}^{-(5)} + \beta_6 \RS_{t-1}^{-(20)} \\
    &+ \beta_7 \Delta J_{t-1}^2,
    \end{split}
\end{align}
with parameters $\beta_0, \ldots, \beta_7 \in \mathbb{R}$. The corresponding HARNets with aggregation periods $(1, 5, 20)$ and $(1, 5, 20, 40, 80)$ are defined by extending the model such that the input to the first convolutional layer is a two-dimensional time series, consisting of the entries for $\RV_t$ and $\RS^-_t$, and each layer has two two-dimensional filters with the same length as in equation \eqref{eq:layer}. Using a similar scheme as introduced in the case of a one-dimensional input time series, the model parameters of these extended HARNets can again be initialized such that they exactly reproduce the predictions of the linear baseline model \eqref{eq:HARRSJ}. Notice that, when using the $\log$-OLS approach, we cannot directly use signed jump variations as inputs since these are not always positive. In this case, as suggested in the appendix to \citet{patton_good_2015}, we replace $\Delta J_{t}^2$ by a percentage
jump variation measure defined as $1 + \Delta J_{t}^2/\RV_{t}$, prior to logarithmizing.

Following the recommendations obtained in Section~\ref{sec:results}, we restrict our numerical experiments in this setting to the case where the QLIKE error is used as a loss function and the NN weights are initialized by the OLS fit of the baseline extended HAR model. Using the same pairs of test and training sets as in Section~\ref{sec:results}, Figure~\ref{fig:additional_base} compares the respectively optimized HARNets to the OLS fit of the baseline HAR model in terms of the MAE, MSE, and QLIKE test error. Again, we find that the extended HARNets can consistently and often substantially improve upon the accuracy of the baseline model with respect to all three test metrics and with larger margins than in our previous results from Section~\ref{sec:qlikegen}.

Figure~\ref{fig:additional_vs} shows an analysis of the performance of QLIKE optimized HARNets with an OLS baseline fit relative to the OLS, WLS, and $\log$-OLS fits of the respective extended baseline models. Similar to our results in Section~\ref{sec:olsbest}, the extended $\HARPhi{20}$ model still outperforms all of the three approaches for fitting the baseline model and matches the performances of the method where the optimal fitting method was selected a-posteriori. However, we also found a slight decrease in the relative performance of the extended $\HARPhi{80}$ model, which is slightly outperformed by the $\log$-OLS fit of the HAR baseline model with the same aggregation periods.
\section{Conclusions}
We proposed a novel NN model, dubbed HARNet, for the forecasting of RV time series. HARNets are based on hierarchies of dilated causal convolutions. Their close relationship to the well-known linear HAR model allows for an explicit initialization scheme after which the initialized HARNet is computationally equivalent to the fitted HAR baseline model. This approach significantly stabilizes the optimization of HARNets and facilitates a comprehensive evaluation of their performance relative to the respective baseline models.

We found that HARNets can substantially improve upon the forecasting accuracy of the respective HAR baseline models. In particular, HARNets that are initialized by OLS fit of the HAR baseline model and optimized by a QLIKE loss function outperform OLS, WLS, and $\log$-OLS fits of the HAR baseline model when considering the daily RV time series as the sole input.

While in practice, NN models are often treated as black boxes, the model parameters of a HARNet can be clearly interpreted. In this respect, we found that the filter weights, which define the features at weekly and monthly time horizons of a HARNet follow a distinct pattern after optimization. In particular, we found that within the previous week, yesterday and the day before yesterday make by far the most significant contributions to today's RV forecast. We suspect that this reflects the fact that single jumps can only be discriminated from extended periods of high RV when looking at least two days in the past. Regarding the monthly time horizon, we found that within the previous month, the importance of single weeks for the RV forecast seems to diminish almost linearly when moving further into the past.

Hierarchies of dilated convolutional layers allow for an exponential growth of the receptive field with the number of model parameters. This suggests that models such as the HARNet should be particularly well-suited for resolving possible long-term dependencies in the data without the necessity of using an overparameterized model. While a HARNet with a receptive field that covers the past four months outperforms the respective HAR baseline model, this advantage disappears when also considering realized semivariances and signed jump variation as additional inputs. In general, our current findings do not suggest that HARNets are consistently better at understanding long-term relationships in a time series than the respective linear baseline models. We think that this issue is a clear candidate for future research and suspect that a well-considered integration of significantly more training data and a diverse set of exogenous input variables could be crucial to succeed in this matter.

Contrary to previous investigations \parencite{poon2020machine}, the QLIKE loss function turned out to be highly applicable for the optimization of HARNets. We collected strong experimental evidence that this can be explained by a stabilizing effect of our proposed initialization scheme on the optimization process. In general, the idea of initializing a complex non-linear model such that it is computationally equivalent to a linear baseline model before optimization is not restricted to RV forecasting and the proposed scheme could easily be adapted for any autoregression task where linear models already yield competitive results. 

A TensorFlow implementation of the HARNet architecture is available on github at \url{https://github.com/mdsunivie/HARNet}.
\section{Acknowledgements}
R.R. gratefully acknowledges support from the Austrian Science Fund (FWF M 2528). The computational results presented have been achieved using the Vienna Scientific Cluster (VSC).
\printbibliography
\end{document}